\documentclass[acmsmall]{acmart}
\AtBeginDocument{%
  }

\setcopyright{acmlicensed}
\copyrightyear{2026}
\acmYear{2026}
\acmDOI{XXXXXXX.XXXXXXX}

\acmJournal{JACM}
\acmVolume{37}
\acmNumber{4}
\usepackage{enumitem}
\usepackage{array}
\usepackage[caption=false,font=normalsize,labelfont=sf,textfont=sf]{subfig}
\usepackage{textcomp}
\usepackage{stfloats}
\usepackage{url}
\usepackage{verbatim}
\usepackage{algorithmic}
\usepackage{graphicx}
\usepackage{xcolor}

\usepackage{balance}

\newcommand{\RankIIPart}[1] {\section{#1}}
\newcommand{\RankIIIPart}[1] {\subsection{#1}}

\newcommand{\RankIIIPartR}[0] {Section}

\newcommand{\FigureWord}[0] {Fig.}
\newcommand{\FigureWords}[0] {Figs.}
\newcommand{\TableWord}[0] {Table}
\newcommand{\EquationWord}[0] {}
\newcommand{\EquationWords}[0] {}

\newcommand{\captionX}[1]{\caption{#1} \Description{#1}}
\newcommand{\sysname}{RateCount}

\begin{document}

\title{\sysname{}: Learning-Free Device Counting by Wi-Fi Probe Listening}

% \author{Tianlang He}
% \email{theaf@cse.ust.hk}
% \orcid{0000-0002-4939-5993}
% \author{Zhangyu Chang}
% \email{zchang@cse.ust.hk}
% \author{S.-H. Gary Chan}
% \email{gchan@cse.ust.hk}
% \affiliation{%
%   \institution{The Hong Kong University of Science and Technology}
%   \city{Hong Kong}
%   \country{China}
% }
\author{Tianlang He}
\orcid{0000-0002-4939-5993}
\affiliation{%
  \institution{The Hong Kong University of Science and Technology}
  \city{Hong Kong}
  \country{Hong Kong S.A.R.}}
\email{theaf@cse.ust.hk}
\author{Zhangyu Chang}
\orcid{0000-0002-1069-4048}
\affiliation{%
  \institution{Yango University}
  \city{Fuzhou}
  \country{China}}
\email{zchang@cse.ust.hk}
\author{Zhongming Lin}
\affiliation{%
  \institution{The Hong Kong University of Science and Technology}
  \city{Hong Kong}
  \country{Hong Kong S.A.R.}}
\email{zmhkust@ust.hk}
\author{S.-H. Gary Chan}
\orcid{0000-0003-4207-764X}
\affiliation{%
  \institution{The Hong Kong University of Science and Technology}
  \city{Hong Kong}
  \country{Hong Kong S.A.R.}}
\email{gchan@cse.ust.hk}

\renewcommand{\shortauthors}{He et al.}

\begin{abstract}
    % Problem and context
Counting Wi-Fi devices within access point (AP) coverage by listening to their probe request frames (PRFs) % with randomized MAC addresses 
is a well-established research problem, fundamental to many Internet of Things (IoT) applications such as crowd management and public transportation scheduling. 
% Gap
While commendable counting accuracy has been reported, existing approaches fall short in deployment convenience due to their reliance on machine learning, which necessitates 1) extensive data collection and training efforts for system setup, and 2) specialized model fine-tuning for operational maintenance. We propose \sysname{}, an accurate, lightweight, and learning-free counting approach to lower deployment costs. \sysname{} employs a provably unbiased closed-form expression to estimate the device count based on the rate at which APs receive PRFs, along with an error model to compute the estimation variance. We also demonstrate its application in people counting by incorporating a device-to-person calibration scheme. Through extensive real-world experiments conducted at multiple sites spanning a wide range of counts, we show that \sysname{}, without any deployment costs for machine learning, achieves comparable counting accuracy to the state-of-the-art (SOTA) learning-based device counting and improves previous people counting schemes by a large margin.

\end{abstract}





\thanks{This work was supported, in part, by Research Grants Council Collaborative Research Fund (under grant number C1045-23G), and RGC-General Research Fund (under grant number 16201625).}

\maketitle

% \begin{IEEEkeywords}\end{IEEEkeywords}

\RankIIPart{Introduction} 
\label{sec:intro}

A Wi-Fi-enabled device, or simply \emph{Wi-Fi device} in this paper, actively manages its wireless connection by sporadically broadcasting probe request frames (PRFs) to the access points (APs) in its proximity. 
The frames are broadcast no matter whether the device has been connected to an AP, with intervals generally ranging from seconds to minutes.
Although each Wi-Fi device has a unique physical MAC address, to protect user privacy, unconnected devices often fabricate random (or virtual) MAC addresses in their PRFs, known as \emph{MAC address randomization}~\cite{freudiger2015talkative, tan2021efficient}.
As a result, an AP may receive multiple distinct virtual MAC addresses from an unconnected device while the device remains in the coverage area. 

As reported in much of the literature, the number of Wi-Fi devices in a {\em counting area} (defined as the joint coverage of one or more APs) may be estimated through listening to their PRFs broadcast within a window (of a few to tens of minutes)~\cite{oppokhonov2022analysis, 9888045, perez2024randomization}. 
Such \emph{device counting} is fundamental to many Internet of Things (IoT) applications, such as crowd estimation, Wi-Fi popularity assessment, and public transportation scheduling~\cite{khan2022crosscount,ushakov2022internet, peng2024single, chen2024metaverse, he2025elevator, zhong2025mtm}. 
The state-of-the-art (SOTA) approaches count devices by the number of MAC addresses. 
To prevent overcounting due to the MAC address randomization, they employ machine learning to associate PRFs with their broadcast devices~\cite{TorkamandiOnlineMethodEstimating2021, NittiiABACUSWiFiBasedAutomatic2020, jin2024over, fenske2021three}. While accurate counting is achieved, their system deployment is often costly. For one thing, the model training for PRF association requires 
massive efforts for collecting and curating training data, and for tuning model hyperparameters~\cite{dong2021survey, zhou2022domain}. For another, the association model needs regular fine-tuning to accommodate new device brands and models~\cite{nozari2024analyzing, fenske2021three}. Hence, deploying these approaches is costly and time-consuming, requiring continuous human efforts, professional expertise, and computing power to support machine learning. 

For fast deployment and easy maintenance, we study the number of PRFs over a window, i.e., {\em probing rate}. Intuitively, whether MAC addresses are randomized or not, APs typically receive more PRFs with more devices in a counting area. 
Thus, we may use probing rates to infer device counts, free from the need for PRF association or machine learning. 
%While this idea has been naively explored, the relationship between probing rate and device count has not been carefully studied. This results in large, unpredictable counting errors. 
While simple, this idea, to the best of our knowledge, has not been systematically studied in the literature, leaving three critical questions open: 1) What is the physical meaning of the device count estimated from the probing rate, and given that meaning, how can it be estimated unbiasedly? 2) How to interpret the counting error and determine the window size? 3) How to extend the device count to people counting?  

\begin{figure}
    \centering
    \includegraphics[width=0.73\linewidth]{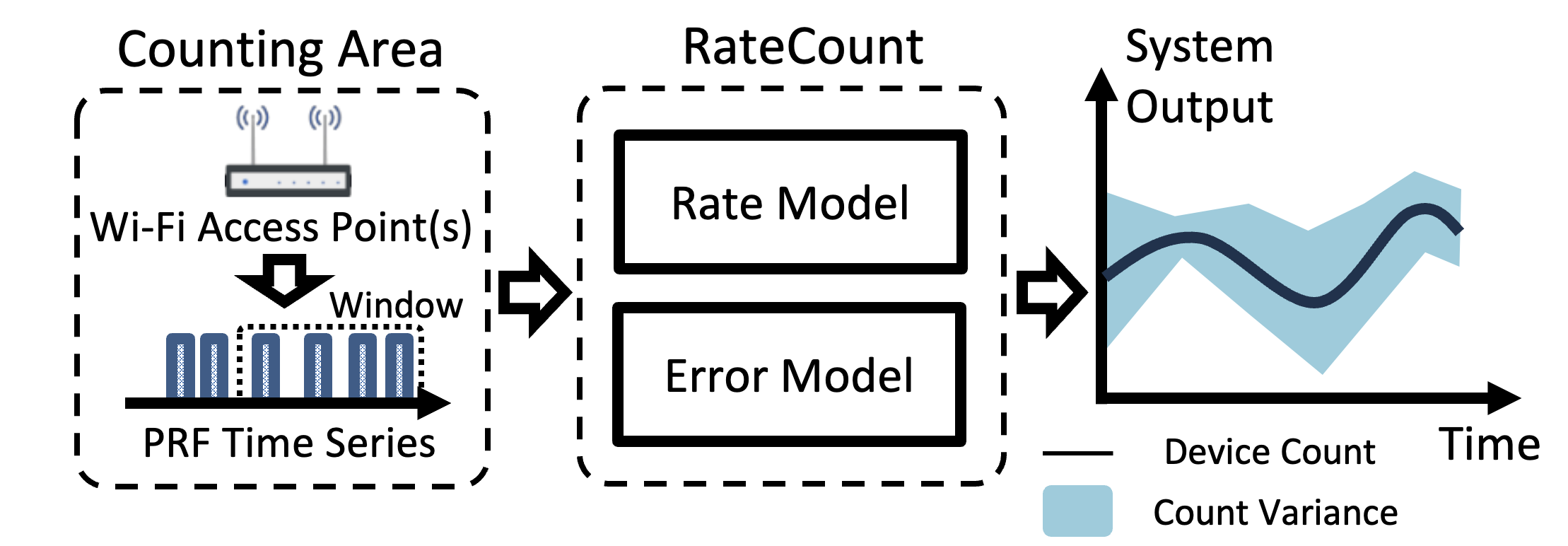}
    \caption{The counting process of \sysname{}. Given a sliding window and a time series of PRFs as listened by APs, \sysname{} uses a rate model to estimate the device count over the window and an error model to compute the estimation variance. }
    \label{fig: sys_diagram}
    \Description{system diagram}
\end{figure}

To answer these questions, we propose \sysname{}, an accurate, lightweight, and learning-free Wi-Fi device counting approach under MAC address randomization. 
\FigureWord~\ref{fig: sys_diagram} shows the counting process of \sysname{}. Given a sliding window and a time series of PRFs as listened by APs, \sysname{} uses two closed-form expressions--a rate model and an error model--to estimate device count and its variance, respectively, and it adapts to environmental changes by itself.   
\sysname{} makes the following contributions: 

\begin{itemize}[leftmargin=*]
    \item \emph{A closed-form rate model to count Wi-Fi devices and its unbiased estimator proof}:
        We propose a rate model--a lightweight, closed-form expression--that estimates device counts by the ratio of the probing rate of all devices within a window to the mean probing rate of a single device. % This achieves accurate device counting with efficient system deployment and lightweight computation. 
        We further prove that our rate model is an unbiased estimator of the device count time-averaged over the window. 
        
    \item \emph{A closed-form error model to estimate the count variance}: 
    We derive a closed-form error model for the variance of the estimated counts, in terms of the window size, the number of PRFs, and the interval variance of device probing. We further provide a thorough analysis of the model to understand the counting error and instruct how to determine window size. 

    \item \emph{Extension to people counting}:
    As a showcase of one major application, we demonstrate how to apply \sysname{} to people counting, based on our deployment experiences from an existing counting system. Specifically, the system employs a multimodal calibration scheme to estimate the device-to-person ratio in the counting area. 
    
\end{itemize}

We have implemented \sysname{} and conducted extensive experiments to count devices and people in various environments, including a lab room, a building entrance, a campus atrium, a study room, and a bus station. Our learning-free \sysname{} shows comparable accuracy with the SOTA learning-based device counting and significantly reduces the counting error of previous Wi-Fi-based people counting by 66\%. Furthermore, our error model demonstrates a sound estimation of the count variance. 

The remainder of the paper is organized as follows.  We first introduce the preliminaries in Section~\ref{sec:preliminary}.  Then, we discuss the design of \sysname{} in Section~\ref{sec:solution}, followed by its extension to people counting in Section~\ref{sec: deployment}.  We present illustrative experimental results in Section~\ref{sec:experiment}.  We finally review related work in Section~\ref{sec:related}, and conclude in Section~\ref{sec:conclusion}. We prove that \sysname{} gives an unbiased estimator in Appendix~A. %~\ref{app: proof}.

\RankIIPart{Preliminaries} 
\label{sec:preliminary}

\begin{table}%[htbp]
  \caption{Summary of Notations (unless specified, all concepts are relative to a specific window in a specific counting area). }
  \label{table:solution_notation}
  \footnotesize
  \renewcommand\arraystretch{1.1}
  \begin{tabular}{lp{6.8cm}}
    \hline
    \textbf{Notation}   & \textbf{Definition} \\
    \hline
    $U$                     & Set of devices $\{u\}$ \\
    $V$                     & Set of counting areas $\{v\}$\\
    $w$                     & Window size \\
    $\mathbf{M}$            & Number of people \\
    $M(t)$                  & Number of people at time instant $t$ \\
    $\overline{M}$          & Average number of people ($\overline{M}= (1/w) \int_0^w M(t) dt$) \\
    $\mathbf{N}$            & Number of devices \\
    $N(t)$                  & Number of devices at time instant $t$ \\
    $\overline{N}$          & Average number of devices ($\overline{N}= (1/w) \int_0^w N(t) dt$) \\
    $\widehat{N}$           & Estimator of $\overline{N}$ \\
    $p_{m}$                 & Dwell time of the $m$th person \\    
    $d_{n}$                 & Dwell time of the $n$th device \\
    $D$                     & Total dwell time of all devices ($D=\sum_{n=1}^\mathbf{N} d_n$) \\ 
    $b_n$                   & Number of probing bursts from the $n$th device\\
    $B$                     & Total number of probing bursts ($B=\sum_{n=1}^\mathbf{N} b_n$) \\
%   $\tau_{i}$              & Duration of $i$th probing interval \\
%   $\overline{\tau}$       & Mean of probing intervals ($\overline{\tau}= (1/b) \sum_b \tau_i$) \\
    $\overline{\tau}$       & Mean of probing intervals ($\overline{\tau} = \int_0^{\infty} \tau Pr(\tau \mid v) d\tau$) \\
    $\sigma_{\tau}$         & Standard deviation of probing intervals \\
    $\overline{r}$          & Mean probing rate of all devices ($\overline{r}=1/\overline{\tau}$) \\
    $R$                     & Aggregated probing rate ($R=B/w$)\\
    $\alpha$                & Device-to-person ratio  ($\alpha=\mathbf{N}/\mathbf{M}$) \\     
%   $\alpha_m$              & Ratio of device-to-person of the $m$th person \\
    \hline
  \end{tabular}
\end{table}

In this section, we introduce the background knowledge of Wi-Fi probing (Section~\ref{subsec:pre_probing}) and device counting (Section~\ref{subsec:pre_counting}). The major symbols are listed in \TableWord~\ref{table:solution_notation}. Unless otherwise specified, all concepts are relative to a specific window size within a specific counting area.

\subsection{Wi-Fi Probing}
\label{subsec:pre_probing}
\begin{figure}
    \centering
    \includegraphics[width=0.4\linewidth]{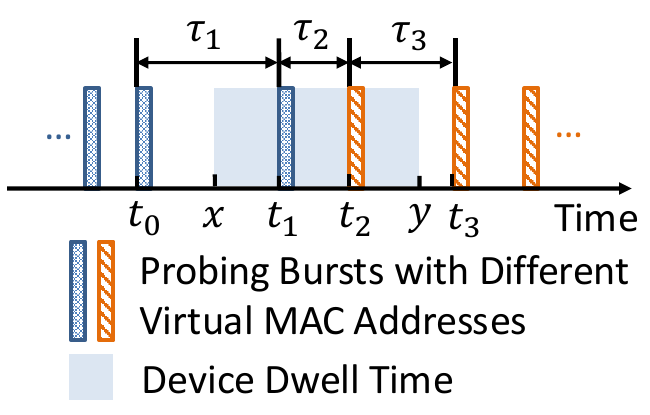}
    \captionX{Probing pattern of a single device and its dwell time in a counting area. 
    The start time of a probing burst is defined as the probing instant, shown as $t$, and the interval between two consecutive probing instants is defined as the probing interval, denoted as $\tau$. The device dwells in a counting area from time $x$ to $y$. }
    \label{fig: probing} 
\end{figure}

\FigureWord~\ref{fig: probing} illustrates three important concepts of Wi-Fi probing that relate to device counting: probing burst, MAC address randomization, and probing interval.  \\

\noindent
\textbf{Probing Burst:} Wi-Fi devices sporadically broadcast PRFs to probe the nearby APs, known as Wi-Fi probing. 
To enhance network discovery, Wi-Fi devices broadcast PRFs in bursts, i.e., device repeats PRF broadcast within a very short period (typically less than 4 seconds), with a much longer interval between two bursts (ranging from 30 seconds to a few minutes)~\cite{freudiger2015talkative}. In \FigureWord~\ref{fig: probing}, the duration of a \emph{probing burst} is presented as a colored bar. We define the start time of a probing burst as the \emph{probing instant}, shown as $t_i$ ($i=0,1,\ldots$), and the interval between two consecutive probing instants as the \emph{probing interval}, denoted as $\tau_j$ ($j=0,1,\ldots$). \\

\noindent
\textbf{MAC Address Randomization:}
At the data-link layer, PRF carries the MAC address of the device. 
Although each Wi-Fi device has a unique physical MAC address, to protect user privacy, most devices irregularly fabricate virtual MAC addresses in their PRFs (when unconnected to an AP), known as MAC address randomization~\cite{fenske2021three}. % \footnote{The occurrence of the randomization process is considered as random and unpredictable~\cite{fenske2021three}. } 
As shown in \FigureWord~\ref{fig: probing}, the probing bursts at $t_1$ and $t_2$ carry different MAC addresses (differentiated by color and hatch) due to the MAC address randomization. Hence, an AP can recognize two bursts from the same device if they share the same MAC address, but two bursts from the same device may not carry the same MAC address.\footnote{Here, we ignore the minimal probability that multiple devices share the same virtual MAC address. } \\

\noindent
\textbf{Distribution of Probing Interval: } 
Probing interval is the idle time of a device before it initiates the next probing burst. Typically, probing interval depends on the device's internal setting (e.g., hardware specifications) and external environment (e.g., signal reception quality); often, it also depends on randomness~\cite{freudiger2015talkative}. Thus, we could interpret the probing interval ($\tau$) as an \emph{independent and identically distributed} (IID) random variable sampled from the statistical distribution that depends on both the device and the external environment. 
Formally, for a single device $u \in U$ in a counting area $v\in V$, the probability that a probing interval is $\tau$ can be expressed as
\begin{equation}
    Pr(\tau \mid u, v), 
    \label{eq: conditional_distribution}
\end{equation}
where $U$ is the set of devices $\{u\}$, and $V$ is the set of counting areas $\{v\}$. 

We will later reveal that knowing the distribution of probing interval is important to achieve unbiased device counting. Unfortunately, knowing the distribution for each device in each counting area is next to impossible, given the unpredictably diverse brands and models of Wi-Fi devices.\footnote{The distribution may differ from device to device, even among devices of the same brand and model. } 
Thus, we have to forgo the device information and collect the interval distribution in each counting area, i.e., the marginal distribution of \EquationWord(\ref{eq: conditional_distribution}). Formally, the probability that a probing interval is $\tau$ in a counting area $v\in V$ is expressed as 
\begin{equation}
    Pr(\tau\mid v)= \sum_{u\in U} Pr(\tau \mid u, v) Pr(u \mid v).  
    \label{eq: marginal_distribution}
\end{equation}
Simply, we can estimate \EquationWord~(\ref{eq: marginal_distribution}) by collecting massive PRFs in a counting area over an extended period. 

\begin{figure}[t]
    \centering
    \begin{minipage}{0.46\linewidth}
        \centering
        \includegraphics[width=\linewidth]{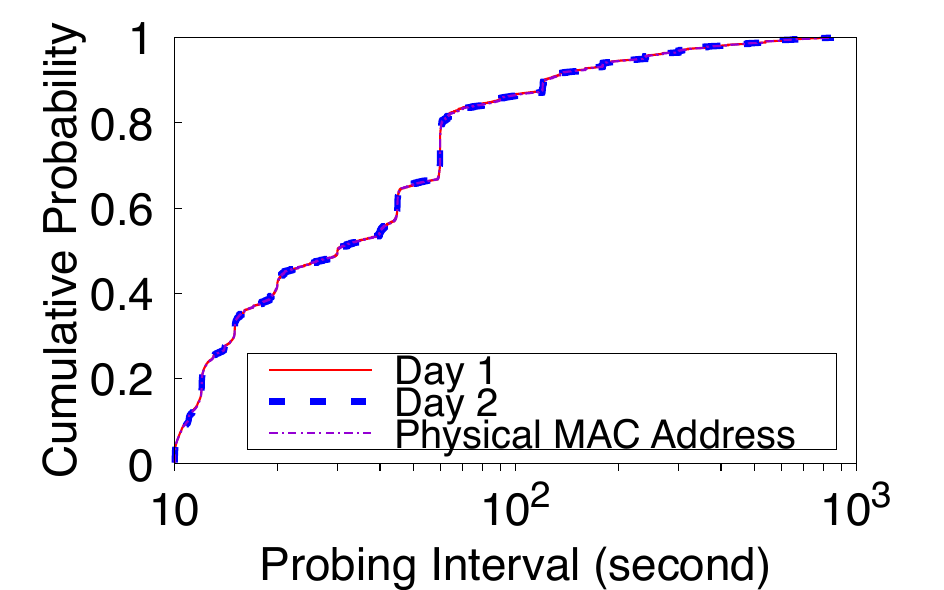}
        \captionX{Distributions of probing interval under different conditions in a counting area.}
        \label{fig:interval_distribution_daily}
    \end{minipage}\hfill
    \begin{minipage}{0.46\linewidth}
        \centering
        \includegraphics[width=\linewidth]{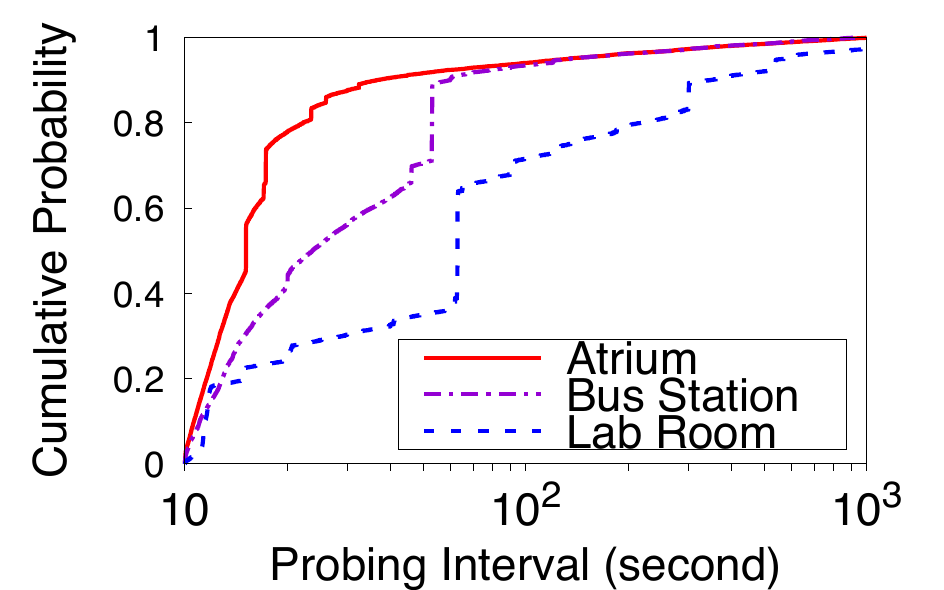}
        \captionX{Distributions of probing interval in different counting areas.}
        \label{fig:culmulative_distribution}
    \end{minipage}
\end{figure}

As a showcase of the data collection, we employ the existing APs from three counting areas (a university atrium, a bus station, and a lab room) for collecting PRFs for multiple days. On average, around 266,000 PRFs were received in each area per day. \FigureWords~\ref{fig:interval_distribution_daily} and~\ref{fig:culmulative_distribution} show the distributions of probing intervals based on a sliding window of a day, with the intervals calculated from those having the same MAC addresses (note that the entire collection process is automatic, without any human effort). 
From the two figures, we can conclude that 
\begin{itemize}[leftmargin=*]
    \item {\em The data collection is sufficient to reflect the area-based distribution}: \FigureWord~\ref{fig:interval_distribution_daily} shows the distributions of one counting area in different days and randomization states (by checking the $7$th bit of the information element in PRFs~\cite{ieeeGuidelinesUseExtended2017}), as the distributions show insignificant discrepancy.\footnote{Quantitatively, we made a hypothesis that probing interval is IID given a counting area. Through \emph{Ljung-Box} test and \emph{Kolmogorov-Smirnov} test to validate the independence and identical distribution, respectively, both tests \emph{failed to reject} the hypothesis with \emph{p-values} $> 0.9$. } This suggests that the data collection is sufficient to reflect the interval distribution of this area.
    \item {\em The data collection is necessary for each counting area: } \FigureWord~\ref{fig:culmulative_distribution} shows the distribution of probing intervals in different counting areas. Due to the different distributions, the interval distribution collected in one counting area cannot be used for another. Hence, the on-site data collection is necessary. 
\end{itemize}

\subsection{Device Counting}
\label{subsec:pre_counting}

Two common definitions of device counting are (1) the number of devices that appear in a window, denoted as $\mathbf{N}$, (2) the number of devices at a specific instant $t$, denoted as $N(t)$. In this paper, we focus on $N(t)$ due to its widespread applications. For example, knowing the $N(t)$ of bus stations benefits the public transportation schedule. Yet, this knowledge is also useful for other crowd management scenarios in reflecting queuing situations (or crowdedness) at public venues, such as canteens, study rooms, banks, hospitals, and washrooms. 

As mentioned, the SOTA approach counts devices by counting MAC addresses. 
While suitable for counting $\mathbf{N}$, it cannot be naively extended to estimate $N(t)$. This is because the probing bursts from different devices are transient and asynchronous. The number of MAC addresses, in nature, cannot indicate the {\em dwell time} of devices, and thus, we cannot know the device count at any instant. To tackle this, the existing expedient assumes that the device dwell time in a counting area equals the probing intervals. However, this could statistically underestimate the device counts, because the actual device dwell time is expected to be longer than the probing intervals. 

\begin{figure}[t]
    \centering
    \includegraphics[width=0.5\linewidth]{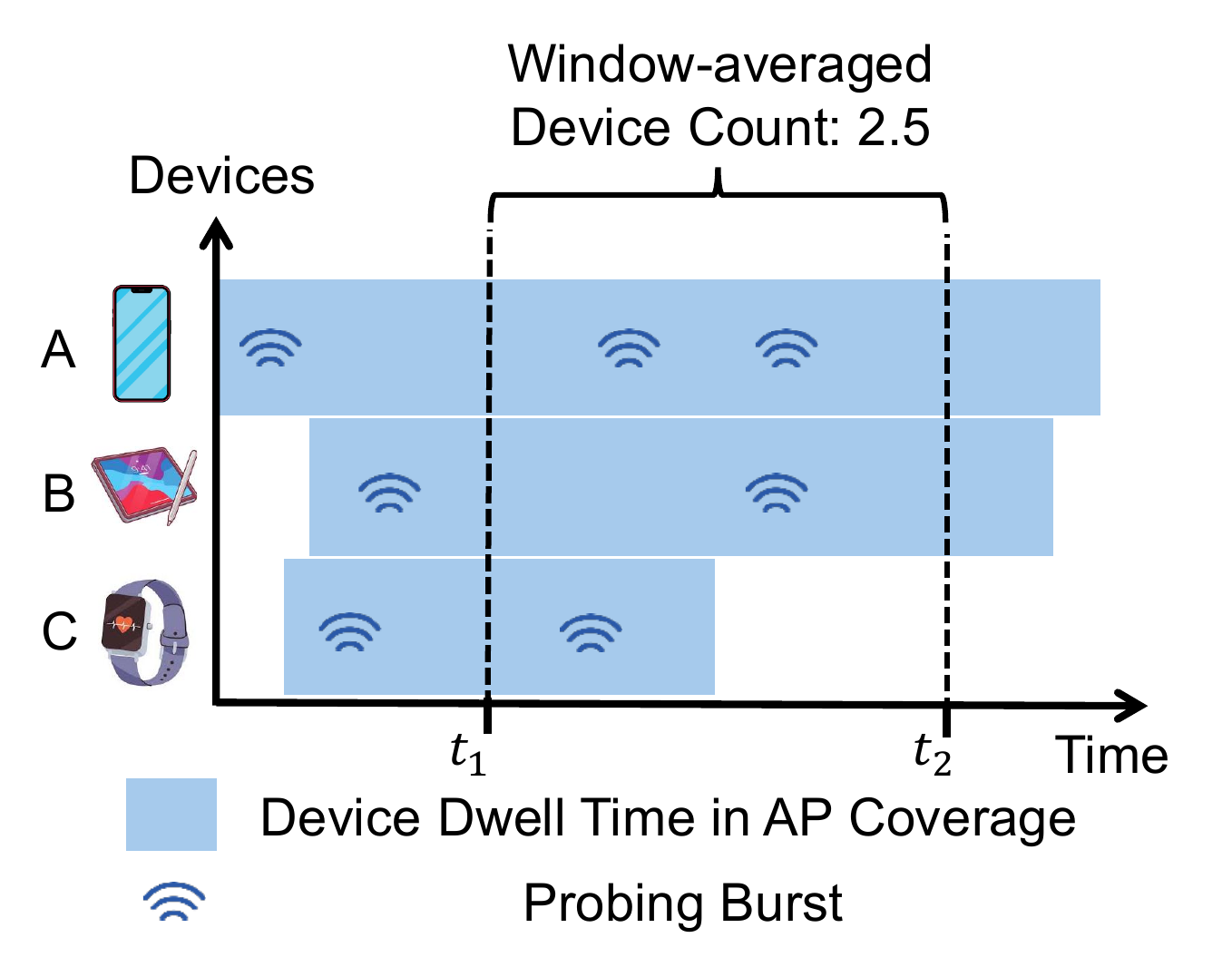}
    \captionX{ Example of device counting (with PRFs already associated with their devices). 
    The window-averaged device count is 2.5, with two devices ($A$ and $B$) fully appearing in the window and one device ($C$) appearing for half of the window. }
    \label{fig: tadc} 
\end{figure}

To substantiate this, we give a counting example in \FigureWord~\ref{fig: tadc}, showing the dwell time and probing bursts of three devices (for simplicity, the probing bursts have already been associated with their devices). 
In this example, due to the short probing bursts, we cannot know the device counts at any instant (say, $t_1$ and $t_2$). 
Even though we assume the probing intervals to be the devices' dwell time, the counts would be statistically underestimated. 
For instance, while the counting at $t_1$ is three, which is accurate (because the probing intervals of $A, B$, and $C$ overlap at $t_1$), 
the two devices at $t_2$ (i.e., $A$ and $B$) would be undercounted since the probing intervals are shorter than the actual dwell time. 

To address this, we propose to approximate $N(t)$ by the number of devices time-averaged over a window of size $w$, termed the window-averaged device count, defined as
\begin{equation}
    \label{eq: mean_device}
    \overline{N}=\frac{1}{w}\int_{0}^w N(t) dt.      
\end{equation}
As illustrated in \FigureWord~\ref{fig: tadc}, the window-averaged device count from $t_1$ to $t_2$ is 2.5, because two devices ($A$ and $B$) appear fully in the window, and one device ($C$) appears for half of the window. One advantage is that this counting objective removes the need for assuming the device dwell time, thus avoiding the underestimation of devices. Furthermore, so long as we find a proper window size ($w$), the window-averaged device count can effectively reflect $N(t)$ with an acceptable delay. 

In the following, we will discuss how to find a window, and compute the unbiased estimator of $\overline{N}$ and the count variance without using machine learning.  

\RankIIPart{\sysname{}: Learning-Free Device Counting}
\label{sec:solution}

In this section, we discuss the design of \sysname{}. We present the rate model for device counting (Section~\ref{subsec: rate_model}), explain its approximation of ground truth (Section~\ref{subsec: proof}), and give a proof of unbiasedness (Appendix~A). Finally, we discuss the error model to estimate counting errors (Section~\ref{subsec: error_model}). 

\subsection{Rate Model}
\label{subsec: rate_model}

Recall that a counting area is the joint coverage of one or more APs. To estimate the device count ($\overline{N}$) in a counting area given a sliding window (of size $w$), we propose a closed-form expression, named the rate model. %Our intuition is simple: since more devices in a counting area lead to an increased rate at which AP receives probing bursts, we can measure the rate to estimate the device count. 
Simply put, the rate model counts devices by calculating the ratio of the aggregated probing rate of all devices to the mean probing rate of a single device. 

%To enable efficient and effective deployment, our rate model aims to fulfill three criterion: 1) it does not require any data collection or training efforts before the system is installed, but it can adapt to the counting area on the fly; 2) it should be an unbiased estimator of the window-averaged device count; 3) its computation should be lightweight. Our rate model includes three steps. 

First, we define the \emph{aggregated probing rate} as the rate at which APs receive probing bursts in a window, which is calculated as
\begin{equation}
    \label{eq: r}
    R=\frac{B}{w},
\end{equation}
where $B$ is the total number of probing bursts received in the window. % In a word, the aggregated probing rate reflects how frequently probing burst occurs in a window. 

Second, we define the \emph{mean probing rate} as the average rate that a single device probes in a counting area, which is calculated as
\begin{equation}
    \overline{r}=\frac{1}{\overline{\tau}}, 
    \label{eq: r_bar}
\end{equation}
where $\overline{\tau}$ is the \emph{mean probing interval} in the counting area. Here, the mean probing interval can be derived from the area-based distribution in \EquationWord~(\ref{eq: marginal_distribution}), which is defined as 
\begin{equation} 
    \overline{\tau} = \int_0^{\infty} \tau Pr(\tau \mid v) d\tau. 
    \label{eq: tau_bar} 
\end{equation}
% In a word, the mean probing rate reflects how frequently a single device probes in this counting area on average. 

Third, the rate model counts devices by the \emph{ratio} of the aggregated probing rate to the mean probing rate, which is given as
\begin{equation}
    \label{eq: n_hat}
    \widehat{N} = \frac{R}{\overline{r}} = \frac{B}{w\overline{r}} =\frac{B\overline{\tau}}{w}.    
\end{equation} 
While simple, the rate model ($\widehat{N}$) is an unbiased estimator of the ground truth ($\overline{N}$), which will be explained next.  
% Next, we explain that $\widehat{N}$ approximates $\overline{N}$ in Section~\ref{subsec: proof}, and give a rigorous proof of the unbiasedness based on the \emph{Elementary Renewal Theorem}~\cite{gallager2013stochastic} in Appendix~A. 

\subsection{Approximation of Ground Truth}
\label{subsec: proof}

% As mentioned, our rate model counts devices by $\widehat{N}=R/\overline{r}=B\overline{\tau}/{w}$. In this subsection, we explain that the rate model ($\widehat{N}$) approximates the ground truth of window-averaged device count ($\overline{N}$). 

\noindent
{\bf Explanation: } The rate model ($\widehat{N}=R/\overline{r}$) intuitively reflects the device counts in a counting area. By definition, the aggregated probing rate ($R$) describes how frequently all devices probe, and the mean probing rate ($\overline{r}$) describes how frequently a single device probes on average.
If there is only one device, $R$ is expected to equal $\overline{r}$, so we have $\mathbb{E} \left(R/\overline{r}\right)=1$. If there are $\bf{N}$ such devices, $R$ is expected to be ${\bf N}$ times as frequent as it is in the single-device scenario, i.e., $\mathbb{E} \left(R\right) = \bf{N}\overline{r}$, such that we have $\mathbb{E} \left(R/\overline{r}\right) = \bf{N}$. 
Here, the explanation is based on intuition, without considering device dwell time~(i.e., $\bf{N}=\overline{N}$). 
In practice, $R$ needs to be measured from a window (of size $w$). Below, we explain that the rate model ($\widehat{N}=B\overline{\tau}/{w}$) approximates the window-averaged device count ($\overline{N}$). 
% Furthermore, we show that the rate model approximates the ground truth, i.e., the window-averaged device count ($\overline{N}$); this is expressed as
% \begin{equation}
%     \label{eq: estimator}
%     \widehat{N} = \frac{R}{\overline{r}} \approx \overline{N}.  
% \end{equation} 
% To interpret this, we need to introduce the concept of device dwell time. 

As mentioned, $\mathbf{N}$ is the total number of devices that appear in a window. Let $d_{n}$ denote the dwell time of the $n$th device, during which it broadcasts $b_n$ bursts ($n=0,1,\ldots, {\bf N}$). The total dwell time of all devices in the window is 
\begin{equation}
    \label{eq:dwell_d}
    D = \sum_{n=1}^{\mathbf{N}}d_{n}. 
\end{equation} 
Recall that $\overline{\tau}$ is the mean probing interval, which estimates the expectation of the probing intervals in a counting area. One device is expected to probe once if it dwells in the area for a duration of $\overline{\tau}$. Conversely, we expect one device to dwell \emph{longer} when it broadcasts \emph{more} probing bursts.\footnote{The device may broadcast zero or multiple bursts during $\overline{\tau}$, but the expected value is the mean across all possibilities. } Therefore, the expectation of the dwell time of the $n$th device is $b_n\overline{\tau}$, which gives us this approximation:
\begin{equation}
    \label{eq:approx}
    d_{n} \approx b_n\overline{\tau}.
\end{equation}  
Extending to the case of $\mathbf{N}$ devices, the total dwell time can be approximated by
\begin{equation}
    \label{eq:dwell_tau}
    D = \sum_{n=1}^{N}d_{n} \approx \sum_{n=1}^N \left(b_n\overline{\tau}\right) = B\overline{\tau},
\end{equation} 
note that the approximation error is negligible for sufficiently large $\mathbf{N}$ and $w$, by the \emph{law of large numbers}. On the other hand, according to the definition of $\overline{N}$ in \EquationWord(\ref{eq: mean_device}), the total dwell time is 
\begin{equation}
    \label{eq:dwell_t}
    D = \int_{0}^{w} N(t)~dt = \overline{N} w.  
\end{equation} 
% recall that $N(t)$ is the number of devices at time instant $t$.   
Combining \EquationWords(\ref{eq:dwell_tau}) and~(\ref{eq:dwell_t}), we have
\begin{equation}
    \label{eq:dwell_eq}
    D = \overline{N} w \approx B \overline{\tau}.
\end{equation}
Recall that $\widehat{N}$ is the rate model, and $\overline{N}$ is the ground truth of the window-averaged device count.  
With \EquationWords(\ref{eq: r}) and (\ref{eq: r_bar}), we rewrite \EquationWord(\ref{eq:dwell_eq}) as
\begin{equation}
    \label{eq:dwell_time}
    \overline{N}  \approx \frac{B/w}{1/\overline{\tau}} = \frac{R}{\overline{r}}=\widehat{N}. 
\end{equation} 
This explains that the rate model ($\widehat{N}$) approximates the window-average device count ($\overline{N}$). \\

\noindent
{\bf Example: } We elaborate on the approximation using an example of a single-device scenario, as shown in \FigureWord~\ref{fig: probing}, which can be extended to the case of multiple devices. In the figure, the APs received $b_n=2$ bursts separately at $t_1$ and $t_2$, and the device enters the area at time $x$ and leaves at $y$. Since the entering and leaving times are largely independent of the probing, $x$ could be any value between $t_0$ and $t_1$, and $y$ could be any value between $t_2$ and $t_3$.\footnote{It should be noted that $x$ cannot be too close to $t_0$ since a burst of PRFs could last for a few seconds, but this is negligible because $\overline{\tau}$ is much larger.} Therefore, the expected dwell time of this device is calculated as 
\begin{equation}
    \begin{aligned}
        \label{eq:dwell_expect_2}
        \mathbb{E}(d_{n})  & = \mathbb{E}(y - x) \\
            & = \frac{1}{t_3-t_2} \int_{t_3}^{t_2} \left( \frac{1}{t_1-t_0} \int_{t_1}^{t_0} (y-x) ~\mathrm{d}x\right)~\mathrm{d}y \\
            & = \frac{t_3+t_2}{2} - \frac{t_1+t_0}{2} = \frac{\tau_1+2\tau_2+\tau_3}{2}.  
    \end{aligned}
\end{equation} 
Since $\tau$ is an IID random variable sampled from the interval distribution in this counting area, $\overline{\tau}$ is the expectation of the variables $\tau_1, \tau_2$ and $\tau_3$. We have $\mathbb{E}(d_{n}) = 2\overline{\tau}=b_n\overline{\tau}$, where $b_n=2$ in this example. \\

\noindent
{\bf Proof:} We further prove that the rate model is an unbiased estimator of the window-averaged device count, which is {\em independent} of the distribution of probing interval. The proof is based on the \emph{Elementary Renewal Theorem}~\cite{gallager2013stochastic}. Interested readers could find it in Appendix~\ref{app: proof}. 

\subsection{Error Model}
\label{subsec: error_model}

We study the mean squared error~(MSE) of the rate model. The error consists of counting variance, denoted as $\mathrm{Var}(\widehat{N})$, and counting bias, denoted as $\mathrm{Bias}(\widehat{N}, \overline{N})$, given as 
\begin{equation*}
\begin{aligned}
    \label{eq:mse}
    \mathrm{MSE}\left(\widehat{N}\right)  & = \mathbb{E}\left[\left(\widehat{N}-\overline{N}\right)^2\right] \\
        & = \mathbb{E}\left[\left(\widehat{N}-\mathbb{E}\left(\widehat{N}\right)\right)^2\right] + \mathbb{E}\left[\left(\mathbb{E}\left(\widehat{N}\right)-\overline{N}\right)^2\right] \\
        & = \mathrm{Var}\left(\widehat{N}\right) + \mathrm{Bias}^2\left(\widehat{N}, \overline{N}\right). 
\end{aligned}
\end{equation*} 
Since the rate model is an unbiased estimator, we neglect the bias term. The MSE can be estimated by counting variance, shown as 
\begin{equation}
    \mathrm{MSE}\left(\widehat{N}\right)\approx \mathrm{Var}\left(\widehat{N}\right).  
\end{equation}
Here, the number of bursts ($B$) and the window size ($w$) are parameters. The count variance depends on the variance of the mean probing interval, shown as 
\begin{equation}
    \mathrm{Var}\left(\widehat{N}\right) = \mathrm{Var}\left(\frac{B\overline{\tau}}{w}\right)=\frac{B^2}{w^2}\mathrm{Var}\left(\overline{\tau} \right).   
    \label{eq: var_N}
\end{equation} 
Note that $\mathrm{Var}(\overline{\tau})$ represents the variance of the mean value over a group of samples with a sample size of $B$ (i.e., the number of bursts). Since $\tau$ is an IID random variable, the count variance depends on the variance of its sample mean, i.e., 
\begin{equation}
    \mathrm{Var}(\overline{\tau})= \frac{\sigma_\tau^2}{B}, 
    \label{eq:var_b}
\end{equation}
where $\sigma_\tau$ is the standard deviation of $\tau$ that is derived from \EquationWord(\ref{eq: marginal_distribution}). 
By combining \EquationWords(\ref{eq: var_N}) and~(\ref{eq:var_b}), the error model is given as
\begin{equation}
        \mathrm{MSE}\left(\widehat{N}\right) \approx 
        \mathrm{Var}\left(\widehat{N}\right) 
        = \frac{B\sigma_\tau^2}{w^2}.   
        \label{eq: mse}
\end{equation}
Overall, the error of the rate model depends on the window size ($w$), the number of bursts in the window ($B$), and the standard deviation of the probing intervals ($\sigma_\tau$). The error model conveys the following messages: 
\begin{itemize}[leftmargin=*]
    \item {\em How to determine the window size: } The window size needs to balance the real-time performance and counting accuracy. Intuitively, we prefer a shorter window to approximate $N(t)$ for lower counting delay. However, according to the error model, a too small window could lead to sparse samples of probing bursts, resulting in large counting errors. Therefore, {\em we should select the largest window as long as the delay is acceptable. }For example, a short window (say, 3 minutes) may be used at a bus stop to promptly reflect the crowd changes, while a longer window (say, 15 minutes) may be suitable for a study room to tune air conditioners.  
    \item {\em Why does probing interval affect counting error: } From the AP perspective, probing bursts can be viewed as the samples of the device dwell time, based on which the rate model estimates device counts. Thus, a larger variance of probing interval leads to larger uncertainty of device counting in the counting area, resulting in errors.   
    \item {\em Why does MSE increase with the number of probing bursts ($B$): } The increase of ground-truth device count concurrently leads to the increase of both $B$ and MSE, not that $B$ leads to MSE. This is supported by the fact that the MSE relative to the ground truth (i.e., NRMSE) decreases with the number of probing bursts, which is shown as: 
    \begin{equation} 
        \mathrm{NRMSE}\left(\widehat{N}\right)=\frac{\sqrt{\mathrm{MSE}\left(\widehat{N}\right)}}{\overline{N}}\approx \frac{\sigma_\tau}{\sqrt{\overline{N}w\overline{\tau}}}=\frac{\sigma_\tau}{\overline{\tau}\sqrt{B}},   
        \label{eq: nrmse} 
    \end{equation} 
note that $\overline{N}\approx B\overline{\tau}/w$ from \EquationWords(\ref{eq: n_hat}). The conclusion is that the absolute counting error {\em increases} with the number of ground-truth devices, while the relative error {\em decreases} with the ground truth. In other words, the counting of the rate model is more accurate in crowded scenarios in terms of relative errors. 
\end{itemize}

    %The counting error increases with the number of bursts from MSE, while the number of bursts does not cause the error to increase. Rather, there is a confounding effect, where an increasing number of devices (i.e., the ground truth) concurrently leads to a rise in the number of bursts and the counting error. To support this, we present normalized root-mean-square error (NRMSE), i.e., the MSE of the rate model relative to the ground truth device count, expressed as 

\section{Application in People Counting}
\label{sec: deployment}

In this section, we demonstrate how to extend \sysname{} to people counting, drawing on our experience with an existing system. We will discuss the system deployment (Section~\ref{subsec:cali}), explain the device-to-person ratio (Section~\ref{subsec:ratio}), and present the error analysis  (Section~\ref{subsec:error_analysis}). 

\subsection{System Deployment}
\label{subsec:cali}

Literature has shown that the number of devices can effectively reflect crowd size~\cite{9888045,Haofinegrainedcrowdanalysis2023}. Yet, some scenarios prefer the more fine-grained people counts. %, whether these people carry Wi-Fi devices or not. 
To fulfill this need, we introduce a calibration scheme that extends \sysname{} to people counting. 

The calibration scheme is intuitive. Device and people counts differ by how many devices a person carries. Thus, we estimate the ratio of the number of devices that a person carries in a counting area, i.e., the \emph{device-to-person} ratio, defined as 
\begin{equation}
    \alpha=\frac{\mathbf{N}}{\mathbf{M}},
\end{equation}
where $\mathbf{M}$ is the number of people that appear in a window. With this ratio, the window-averaged people count in the counting area is estimated as  
\begin{equation}
\begin{aligned}
        \overline{M}&=\frac{1}{w}\int_0^w M(t)dt \\
        &\approx \frac{1}{w}\int_0^w \frac{N(t)}{\alpha}dt\approx\frac{\widehat{N}}{\alpha}, \\ 
\end{aligned}
\label{eq:count_time}
\end{equation}
Here, $M(t)$ is the number of people at time instant $t$, and recall that $\widehat{N}$ is the rate model. 

The device-to-person ratio is shown to be highly related to counting scenarios~\cite{WuCrowdEstimatorApproximatingcrowd2018,hao2024heterogeneous,solmaz2022countmein}. 
For example, in an elderly home, the ratio is usually less than $1$ because many residents do not use Wi-Fi devices. In contrast, the ratio at a university is often greater than $1$, since most students have more than one device (e.g., smartphones, wristbands, laptops). Therefore, to estimate the ratio in a counting area, we may calculate the ratio from the same scenario in the vicinity. For example, we may calculate the device-to-person ratio at a campus entrance and apply the ratio to a classroom, leveraging the people-sensing facilities (such as a surveillance camera) that campus entrance usually installs. There, we refer to the sensing area of people sensing in the vinicity as the {\em calibration region}. 

\begin{figure}
    \centering
    \includegraphics[width=0.65\linewidth]{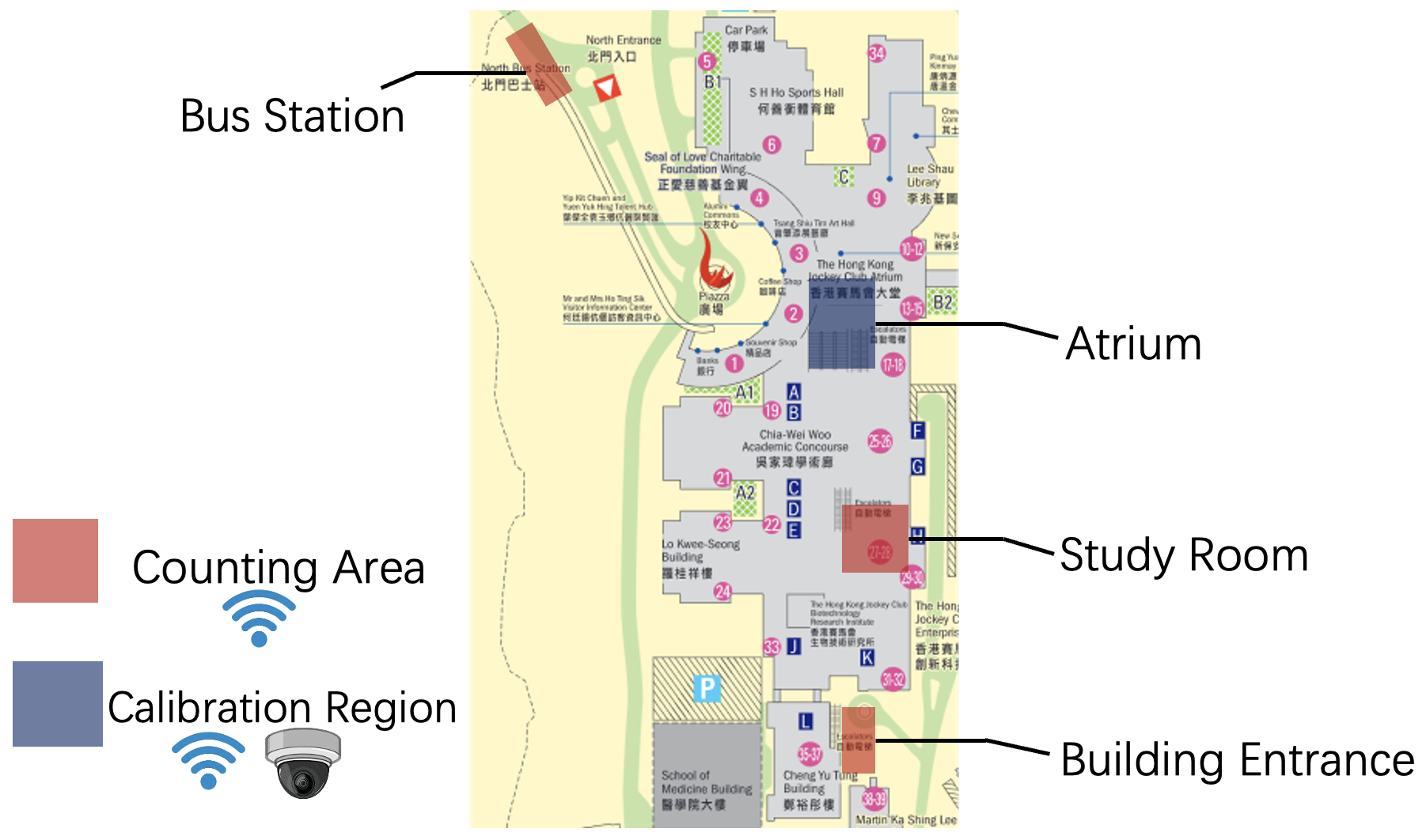}
    \captionX{System deployment of people counting in a university campus. The device-to-person ratio is estimated from a multimodal calibration region and is used for people counting in other areas. }
    \label{fig: calibration_region}
\end{figure}

\begin{figure}[t]
    \centering
    \begin{minipage}{0.49\linewidth}
        \centering
        \includegraphics[width=0.8\linewidth]{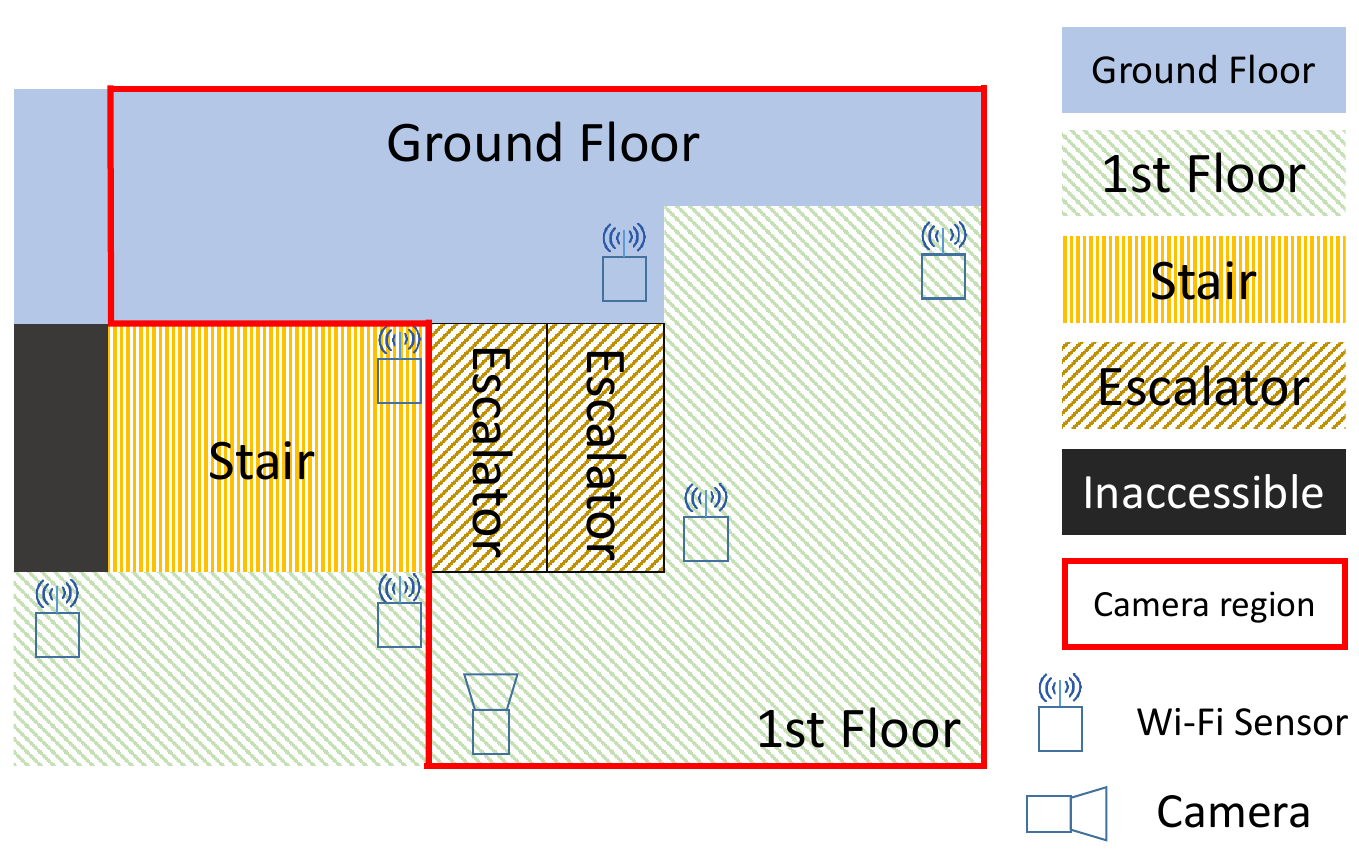}
        \captionX{Sensor deployment in calibration region. The infrastructural APs and cameras are used to estimate the device-to-person ratio. }
        \label{fig:exp_setting}
    \end{minipage}    \hfill
    \begin{minipage}{0.49\linewidth}
        \centering
        \includegraphics[width=0.75\linewidth]{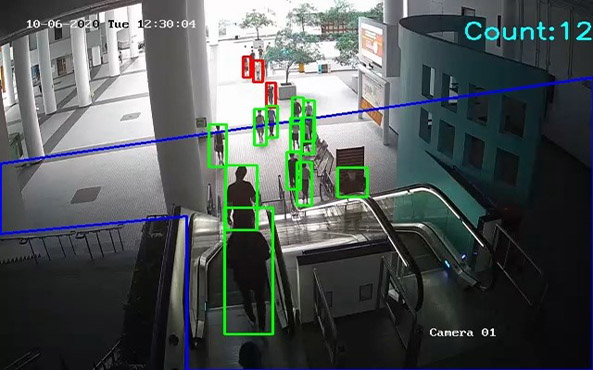}
        \captionX{Example of system operation in calibration region. }
        \label{fig:yolo}
    \end{minipage}
\end{figure}

\FigureWord~\ref{fig: calibration_region} illustrates our existing system deployment in a university campus. The calibration region is in an atrium, and we want to count people at a bus station, a study room, and a building entrance. \FigureWord~\ref{fig:exp_setting} shows the sensor deployment in the calibration region, where $6$ APs and $1$ camera are used to estimate the device-to-person ratio. Specifically, we use \sysname{} to compute $\widehat{N}$ and the object detection~\cite{wang2023yolov7} to estimate the window-averaged people count, denoted as $\widehat{M}$ (an example of estimating $\widehat{M}$ is shown in \FigureWord~\ref{fig:yolo}). Then, we estimate the device-to-person ratio by 
\begin{equation}
    \alpha\approx \frac{\widehat{N}}{\widehat{M}}. 
\end{equation}
The approximation will be explained later. 

For the multimodal calibration, the sensing area of an AP is often wider than that of a camera. To avoid the spill-over of Wi-Fi signals, we only count the probing bursts from the devices within the camera's counting area. Specifically, the probing bursts are counted by  
\begin{equation}
    B'=\sum_{i=1}^{B}Pr(u_i\in v), 
\end{equation}
where $v$ is the counting area of camera, and $u_i$ is the device that broadcasts the $i$th burst. In our deployment, we employ SVM to estimate $Pr(u_i\in v)$ using the RSSI of PRFs (more details of such {\em geofencing technology} can be found in~\cite{tan2020iot}). This technology could also apply to the Wi-Fi counting areas that require a clear semantic boundary. 

\subsection{Device-to-Person Ratio}
\label{subsec:ratio}

As mentioned earlier, the device-to-person ratio is defined as $\alpha=\mathbf{N}/\mathbf{M}$, while in system deployment, we estimate the ratio by $\alpha\approx \widehat{N}/\widehat{M}$. We have shown that $\widehat{N}\approx \overline{N}$, and it is intuitive that $\widehat{M}\approx\overline{M}$. Thus, to justify the system deployment, we explain that 
\begin{equation}
    \alpha\approx \frac{\overline{N}}{\overline{M}}.  
    \label{eq: alpha_approx}
\end{equation}
 
% Next, we explain \EquationWord(\ref{eq: alpha_approx}) by considering the dwell time of devices and people. 
Recall that the total dwell time of all devices is the sum of the dwell time of each device.  
On the other hand, the dwell time of devices equals their users' dwell time.\footnote{Devices that are not associated with any users should be excluded using the method in~\cite{tan2021efficient}} Thus, the dwell time of all devices is given as 
\begin{equation}
    D=\sum_{n=1}^\mathbf{N} d_n = \sum_{m=1}^\mathbf{M} d_m\alpha_m, 
    \label{eq: d_to_p_dwell}
\end{equation}
where $d_m$ is the dwell time of the $m$th person who has $\alpha_m$ as the number of devices in total.\footnote{Rigorously, $\alpha_m$ represents the device-to-person ratio. Since it is specific to one person (the $m$th person), its value can be interpreted as the number of devices that the person carries.} Here, we consider a general case where a person's dwell time is independent of how many devices it carries. 
With a sufficiently large number of people, 
we have
\begin{equation}
    \sum_{m=1}^\mathbf{M} d_m\alpha_m \approx \frac{\mathbf{N}}{\mathbf{M}}\sum_{m=1}^\mathbf{M} d_m,
    \label{eq: p_independent}
\end{equation}
recalling that $\mathbf{N}/\mathbf{M}$ is the mean number of devices carried by a person. 
By combining \EquationWords(\ref{eq: d_to_p_dwell}) and (\ref{eq: p_independent}), 
we get 
\begin{equation}
    \frac{\overline{N}}{\overline{M}}=\frac{\frac1w\sum_{n=1}^\mathbf{N} d_n}{\frac1w\sum_{m=1}^\mathbf{M} p_m}
     \approx \frac{\frac{\mathbf{N}}{\mathbf{M}}\sum_{m=1}^\mathbf{M} p_m}{\sum_{m=1}^\mathbf{M} p_m} = \alpha.  
\end{equation}
This explains that $\alpha$ can be estimated by \EquationWord(\ref{eq: alpha_approx}). 
From the analysis, the calibration region is preferably set up in an area with large crowds, such as a campus atrium or the entrance of a shopping mall. 
Also, a large window size is preferred in the calibration region. 

\subsection{Error Analysis}
\label{subsec:error_analysis}

The error of people counting is similar to that of the device counting. The difference is that we need to consider the error incurred in the calibration scheme. Let $\widehat{M}_c$ and $\widehat{N}_c$ denote the people count and device count estimated in the calibration region, respectively, and recall that $\widehat{N}$ denotes the device count estimated in the counting area.
% The people count in the counting area, denoted by $\widehat{M}$, is given as
% \begin{equation}
%     \label{eq: people_count}
%     \widehat{M} = \frac{\widehat{M}_c}{\widehat{N}_c} \widehat{N}.    
% \end{equation}
The error of people counting is estimated by the error of device counting in the counting area and the propagation of the estimation errors of the two modalities in the calibration region, which is given as  
% Their estimation errors in a calibration region propagate to the NRMSE of the people counting~\cite{taylor1980introduction}, given as 
% According to error propagation, the NRMSE of the people count estimation, denoted by $\mathrm{NRMSE}{(\widehat{M})}$, is given as
\begin{equation*} 
\mathrm{NRMSE}{\left(\widehat{M}\right)}=\sqrt{\mathrm{NRMSE}^2{\left(\widehat{M}_c\right)} + \mathrm{NRMSE}^2{\left(\widehat{N}_c\right)} + \mathrm{NRMSE}^2{\left(\widehat{N}\right)}}.  
\label{eq: people_count_error} 
\end{equation*}
%where the counting error depends on the errors of the device counting in the counting area and the errors of the two modalities in the calibration region. 
Here, we estimate the NRMSE of device counting using the error model in \EquationWord~(\ref{eq: nrmse}). The error of people counting may be subject to the sensing technologies. In our experiment, we use camera-based people counting, and its error ($\mathrm{NRMSE}(\widehat{M}_c)$) is empirically estimated.

\iffalse

Given the popularity of Wi-Fi devices today, the device count is often used to estimate the crowds. In some scenarios, it is preferable to know the exact number of people, whether they have Wi-Fi devices or not. To address this need, we demonstrate a calibration scheme that extends \sysname{} to counting the number of people. 

\fi
\RankIIPart{Illustrative Experimental Results}
\label{sec:experiment}

In this section, we show implementations and extensive experiments to validate \sysname{}. We discuss the experimental setting (\RankIIIPartR~\ref{subsec:environment}), and present experimental results of device counting (\RankIIIPartR~\ref{subsec:result_device}) and people counting (\RankIIIPartR~\ref{subsec:result_people}).

\RankIIIPart{Experimental Setup and Performance Metrics}
\label{subsec:environment}

\begin{figure}[t]
    \centering
    \begin{minipage}{0.48\linewidth}
        \centering
        \includegraphics[width=\linewidth]{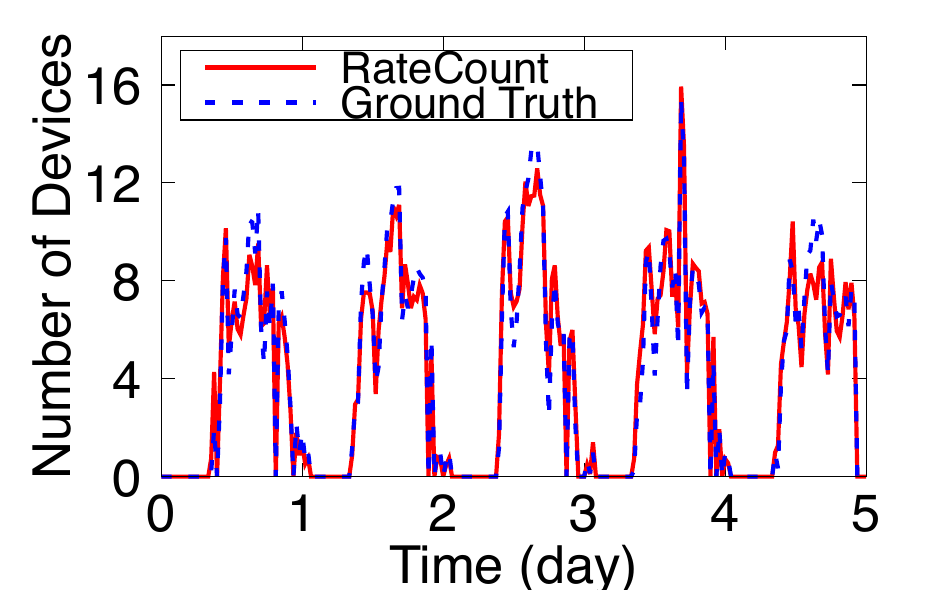}
        \captionX{Device counting in a lab room.}
        \label{fig:people_count}
    \end{minipage}\hfill
    \begin{minipage}{0.48\linewidth}
        \centering
        \includegraphics[width=\linewidth]{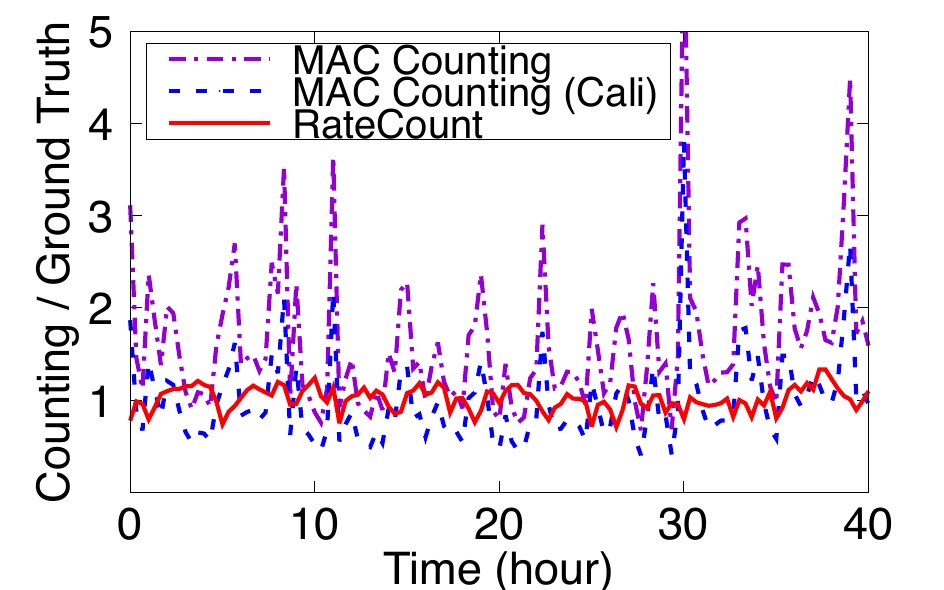}
        \captionX{Comparison between \sysname{} and the MAC-counting approach.}
        \label{fig:correlation}
    \end{minipage}
\end{figure}

Our experiments include device counting and people counting in different subareas of a university campus. To validate device counting, we implement \sysname{} in a laboratory room, where we manually track the entering and leaving times of each device to provide ground truth. To validate people counting, we set up a calibration region in an atrium and count people at a bus station, a building entrance, and a study room, as mentioned in Section~\ref{subsec:cali}. At each site, we employ the infrastructural Wi-Fi APs (GL-AR150 with OpenWrt 18.06) to capture PRFs based on Wi-Fi sniffers implemented with libcap. Once received, the PRFs are transmitted to a personal computer (with an Intel Core i7-4770 3.6 GHz CPU and 32 GB RAM) that runs \sysname{}. To obtain the near-ground truth of people counting, we temporarily install cameras in those sites. Note that the system hardware was chosen empirically, and our approach is independent of this choice. 

We compare \sysname{} with the following device counting and people counting approaches:
\begin{itemize}[leftmargin=*]
    \item \emph{MAC Counting}~\cite{oppokhonov2022analysis} is a classic device-counting method that counts devices by the number of unique MAC addresses. 
    In the experiment, we implement the method by computing the counts averaged over a window. To account for the MAC address randomization, we apply a ratio to calibrate the counts, termed {\em MAC Counting (Cali)}. 
    \item \emph{Learning-based Counting}~\cite{9888045} is the state-of-the-art (SOTA) device-counting approach that uses machine learning to associate the PRFs from the same devices. In the experiment, we implement the method based on the association results over a window. 
    \item \emph{CountMeIn}~\cite{solmaz2022countmein} is the SOTA probing-based people-counting approach that calibrates the number of MAC addresses in a window to people counts using a camera-based calibration scheme.  
\end{itemize}  

We evaluate our approach in terms of window-averaged device counting. Let $\widehat{y}$ represent a counting result, $\overline{y}$ represent the ground truth, and $J$ represent the number of observations in the experiment. We use the following metrics:  
\begin{itemize}[leftmargin=*]
    \item \emph{Root-Mean-Square Error} (RMSE): 
    % which is given as
    \begin{equation}
        \label{eq:rmse2}
        \mathrm{RMSE}\left(\widehat{y}\right) = \sqrt{\frac1J\sum_{j=1}^{n}\left(\widehat{y}_j-\overline{y}_j\right)^2}.
    \end{equation} 
    \item \emph{Mean Absolute Percentage Error} (MAPE): % , which is given as
    \begin{equation}
        \label{eq:mape}
        \mathrm{MAPE}\left(\widehat{y}\right) = \frac{1}{J}\sum_{j=1}^{J}\frac{\left|\widehat{y}_j-\overline{y}_j\right|}{\overline{y}_j}.
    \end{equation} 
    \item \emph{Normalized Root-Mean-Square Error} (NRMSE): %, which is given as
    \begin{equation}
        \label{eq:nrmse2}
        \mathrm{NRMSE}(\widehat{y}) = \frac{\mathrm{RMSE}(\widehat{y})}{\frac1J\sum_{j=1}^J \overline{y}_j}.
        % \frac1n\sum_{i=1}^{n}\frac{\sqrt{(\widehat{y}_i-\overline{y}_i)^2}}{\overline{y}_i}.
    \end{equation}
\end{itemize}
Specifically, RMSE calculates the counting error in terms of the number of devices or people, and MAPE and NRMSE reflect the error relative to the ground truth. It is worth noting that RMSE and NRMSE calculate errors based on the second-order norm, which is more sensitive to large errors than MAPE.  

% The whole system is implemented in Python, and the Wi-Fi sniffer for listening to PRFs is built on \emph{libpcap}. 
% Due to privacy regulations, the camera operated for five hours in the atrium and building entrance, and one hour at the bus station (during a peak hour). 
Unless otherwise specified, the distributions of probing intervals were obtained using a 1-day sliding window (examples are shown in \FigureWord~\ref{fig:culmulative_distribution}). We use a 15-minute window for counting in the lab room (since users there often stay for extended periods) and a 3-minute window in other sites. The step size of the sliding window is 3 minutes.

\RankIIIPart{Device Counting}
\label{subsec:result_device}

\begin{figure}[t]
    \centering
    \begin{minipage}{0.48\linewidth}
        \centering
        \includegraphics[width=\linewidth]{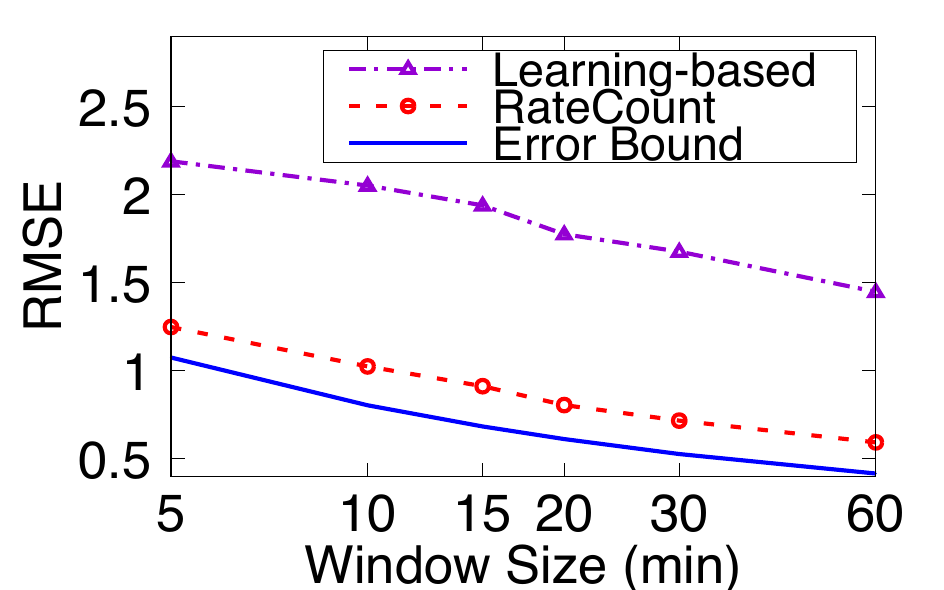}
        \captionX{\sysname{} achieves comparable accuracy with the SOTA learning-based approach.}
        \label{fig:error_WIPEC}
    \end{minipage}\hfill
    \begin{minipage}{0.48\linewidth}
        \centering
        \includegraphics[width=\linewidth]{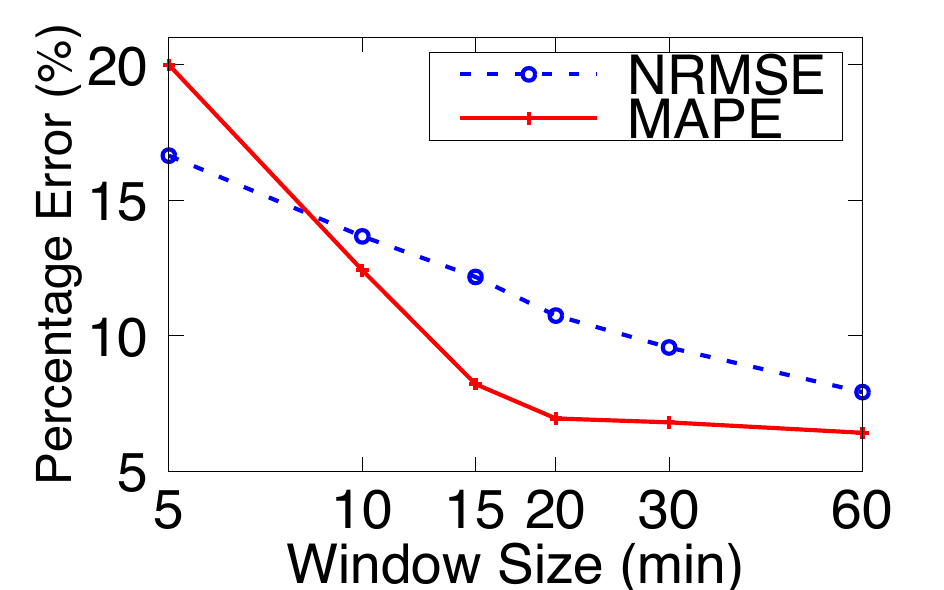}
        \captionX{Error analysis in device counting.}
        \label{fig:relative_error_0}
    \end{minipage}
\end{figure}

\FigureWord~\ref{fig:people_count} shows the results of device counting in a lab room over five days. Therein, \sysname{} accurately tracks the ground truth. 
In particular, the counting results effectively reflect the usage pattern of the lab room: most lab users arrive in the morning and leave in the evening, with the number of devices peaking at noon. This result, using only the infrastructural Wi-Fi APs without any sensor installation or learning from datasets, has been used to tune the air conditioning in the lab room, effectively. 

\begin{figure}[t]
    \centering
    \begin{minipage}{0.48\linewidth}
        \centering
        \includegraphics[width=\linewidth]{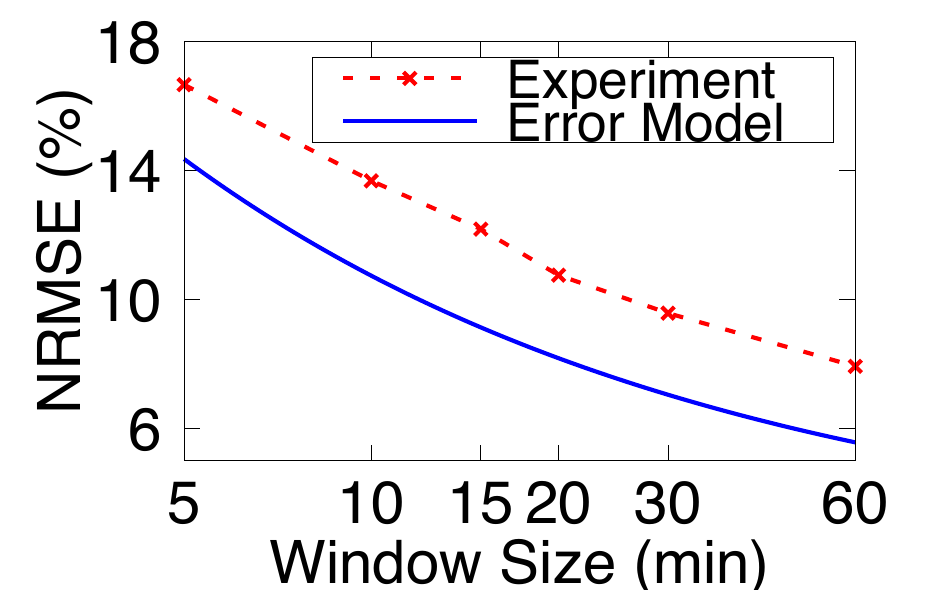}
        \captionX{Comparison between error model and experimental errors in device counting.}
        \label{fig:error_model_0}
    \end{minipage}\hfill
    \begin{minipage}{0.48\linewidth}
        \centering
        \includegraphics[width=\linewidth]{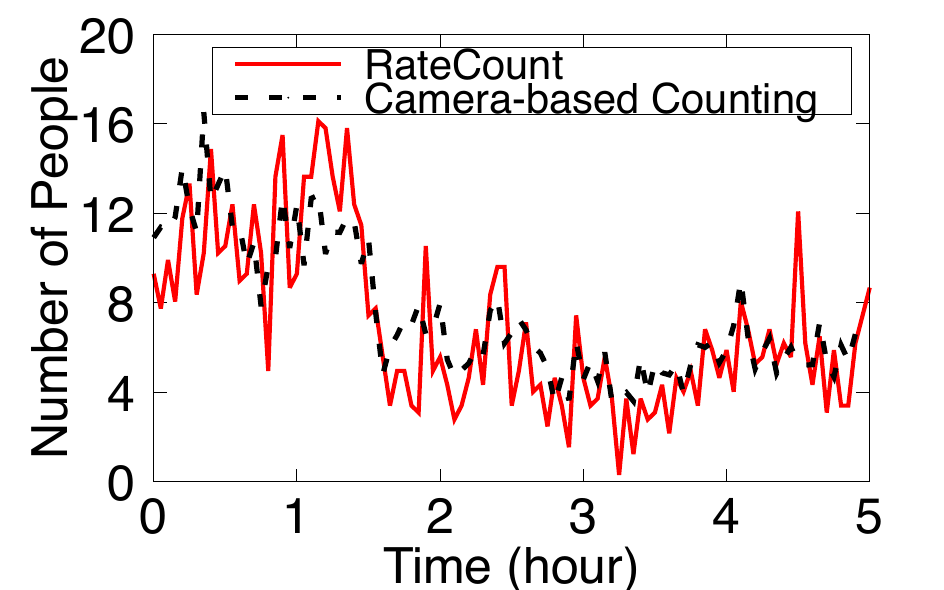}
        \captionX{People counting at a campus atrium.}
        \label{fig:crowd_count_atrium_3min}
    \end{minipage}
\end{figure}

In \FigureWord~\ref{fig:correlation}, we compare \sysname{} with the classic MAC counting method in device counting, showing the counting results over the ground truth (the cases of divided-by-zero have been removed). While both approaches are free from machine learning, \sysname{} outperforms MAC counting by a large margin. This is because the probing rate is independent of MAC address randomization, making it more stable than the number of MAC addresses for device counts. Furthermore, simply calibrating the MAC counts cannot address the randomization issue (here, we empirically set the calibration ratio to $0.5$, though other ratios lead to the same conclusion). This validates that achieving accurate learning-free device counting is nontrivial. 

In \FigureWord~\ref{fig:error_WIPEC}, we compare \sysname{} with the SOTA learning-based device counting, showing RMSE versus window size. 
Overall, the counting accuracy grows with window size, which agrees with our error model that a larger window leads to smaller counting errors (see \EquationWord~(\ref {eq: mse})). In the figure,  
\sysname{} achieves comparable (or even slightly better) results than the learning-based approach, without any deployment costs for machine learning. For one thing, \sysname{} is free from the PRF association that learning-based approaches necessitate, which might lead to minor errors (subject to the effectiveness of the association model). For another, the rate model of \sysname{} is an unbiased estimator of the device counts, which is free from the underestimation issue, as discussed in Section~\ref{subsec:pre_counting}. 

In \FigureWord~\ref{fig:relative_error_0}, we present an error analysis of \sysname{}. As mentioned, the error reduces with an increasing window size, as a larger window leads to lower count variance. Furthermore, the error is more evenly distributed in NRMSE than in MAPE. This indicates that the counting is accurate most of the time, while in very rare cases, there are large errors due to drastic changes in the number of devices. 

\begin{figure}[t]
    \centering
    \begin{minipage}{0.48\linewidth}
        \centering
        \includegraphics[width=\linewidth]{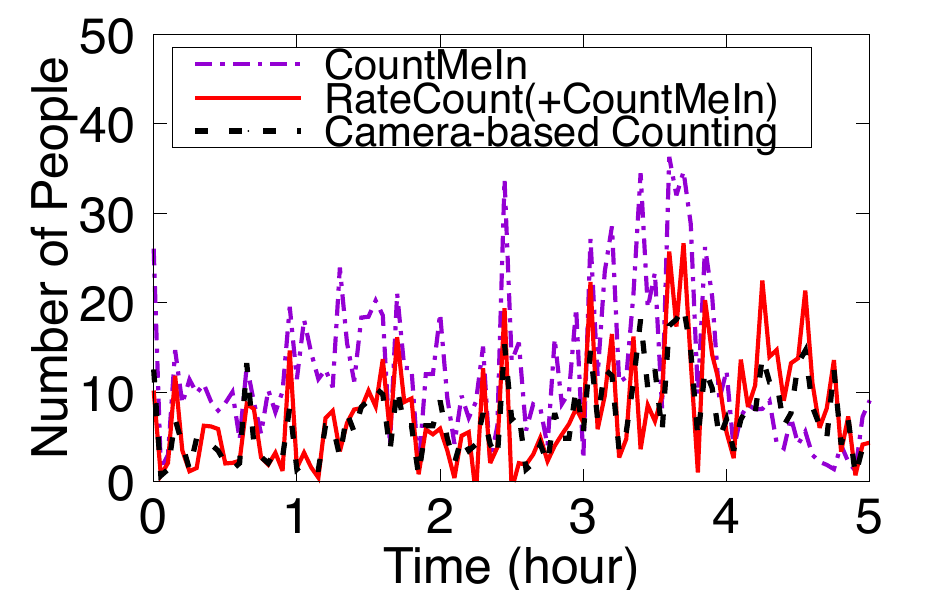}
        \captionX{Evidence that \sysname{} improves the SOTA Wi-Fi-based people counting scheme, demonstrated at a building entrance.}
        \label{fig:crowd_count_cyt}
    \end{minipage}\hfill
    \begin{minipage}{0.48\linewidth}
        \centering
        \includegraphics[width=\linewidth]{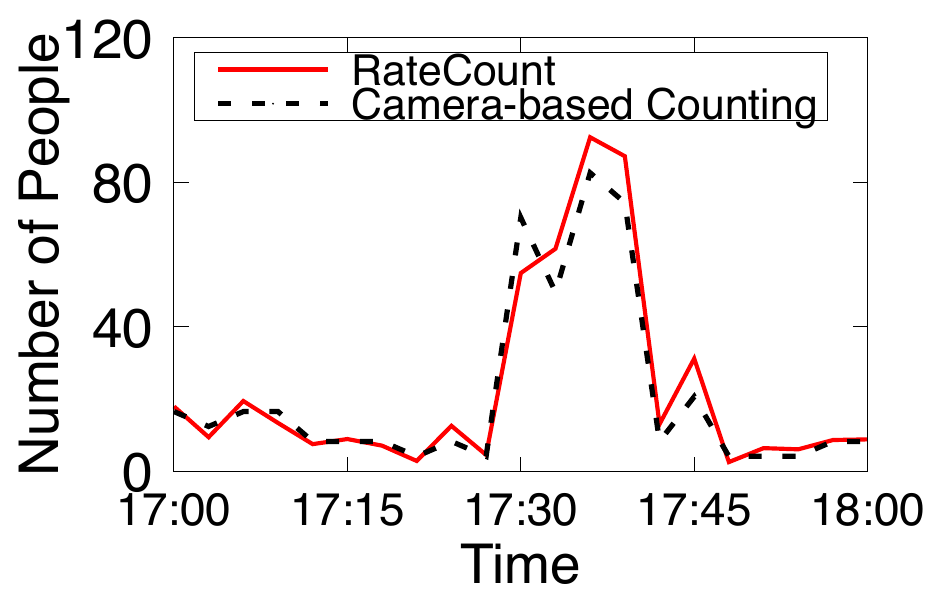}
        \captionX{People counting at a bus station during rush hour.}
        \label{fig:crowd_count_bus_station}
    \end{minipage}
\end{figure}

In \FigureWord~\ref{fig:error_model_0}, we compare the error model with the experimental errors. 
In the figure, the error model not only outlines the error lower bound but also accurately estimates the counting error. Specifically, the experimental errors are larger than the error model by 2\% in NRMSE, which is negligible (say, if there are $10$ devices in total, the gap is $0.2$ devices). This has validated the applicability of our error model in practice.  

\RankIIIPart{People Counting}
\label{subsec:result_people}

\begin{figure}[t]
    \centering
    \begin{minipage}{0.48\linewidth}
        \centering
        \includegraphics[width=\linewidth]{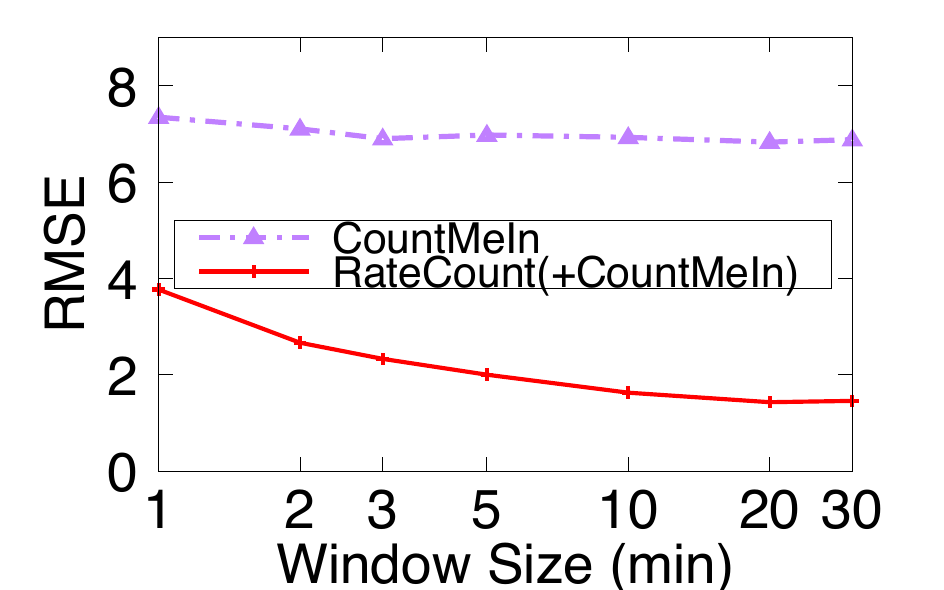}
        \captionX{Error of people counting at a building entrance.}
        \label{fig:relative_error}
    \end{minipage}\hfill
    \begin{minipage}{0.48\linewidth}
        \centering
        \includegraphics[width=\linewidth]{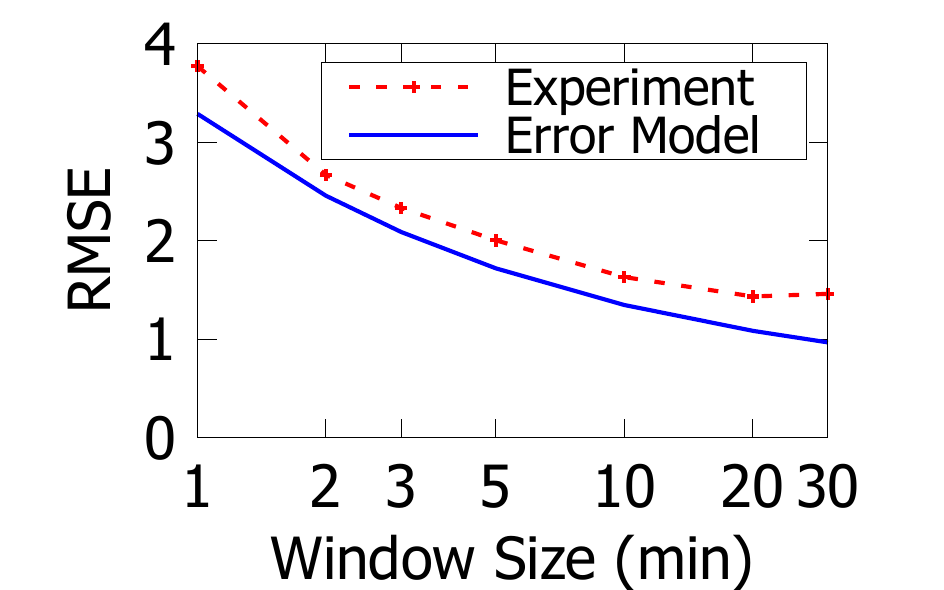}
        \captionX{Comparison between error model and experimental errors in people counting.}
        \label{fig:error_model_1}
    \end{minipage}
\end{figure}

In our experiment of people counting, the device-to-person ratios of the three sites (i.e., the atrium, bus station, and building entrance) are similar (around 1.14), because these sites share a similar scenario, as mentioned in Section~\ref{subsec:cali}. \FigureWord~\ref{fig:crowd_count_atrium_3min} presents the people counting of \sysname{} in the atrium. There, our privacy-preserving Wi-Fi-based approach accurately tracks the outputs of the camera-based people counting, showing comparable counting accuracy with the camera-based approach.  
Note that the camera-based counting approach only provides the near-ground truths, which might not be entirely accurate. In our experiment, we observe that the error of the camera-based people counting is around 0.08 in terms of NRMSE. 

In \FigureWord~\ref{fig:crowd_count_cyt}, we compare \sysname{} with the SOTA probing-based people counting approach (i.e., CountMeIn). The difference of the two approaches is that CountMeIn calibrates MAC counting to the people counts, while ours is based on \sysname{}. In the figure, the counts of CountMeIn fluctuate significantly, since the fluctuating counts of the MAC-counting approach propagate to people counting (due to the MAC address randomization). In comparison, our people counting achieves much lower counting variance owing to the stable and accurate device counting of the rate model. 

\FigureWord~\ref{fig:crowd_count_bus_station} shows the performance of \sysname{} at a bus station during a peak hour. This site is more crowded compared with the atrium scenario. In the figure, \sysname{} shows consistent results with camera-based people counting, effectively capturing the usage pattern that students went to the bus station after the last class concluded at 5:30 PM. 
Such information is important for planning the bus schedule, acquired in a privacy-preserving way.   
Overall, this experiment validates \sysname{}'s scalability to high crowd levels.   

\FigureWord~\ref{fig:relative_error} shows the RMSE of CountMeIn and \sysname{} at a building entrance, using camera-based people counting as the near-ground truth. In the figure, the RMSE of both approaches reduces with the window size, which is as expected. However, the RMSE of \sysname{} is not only much lower than that of CountMeIn, but it also reduces faster as the window size increases. This is because \sysname{} is based on probing rate, free from the perturbation of the MAC address randomization. Thus, \sysname{} shows more similar results to the camera-based people counting. 
Quantitatively, \sysname{} reduces around 4.57 RMSE (around 0.66 NRMSE) from the SOTA Wi-Fi-based people-counting approach. 

\FigureWord~\ref{fig:error_model_1} shows an error analysis of people counting in the three sites (the atrium, building entrance, and bus stop). 
Compared with that of the device counting, the error model of people counting additionally includes the estimation of the people sensing in the calibration region. 
Thus, the result is consistent with the device counting, where the model satisfactorily reflects the actual errors. 
This validates that our error model can also effectively provide a sound estimate of the error in people-counting systems. 

\begin{figure}
    \centering
    \includegraphics[width=0.85\linewidth]{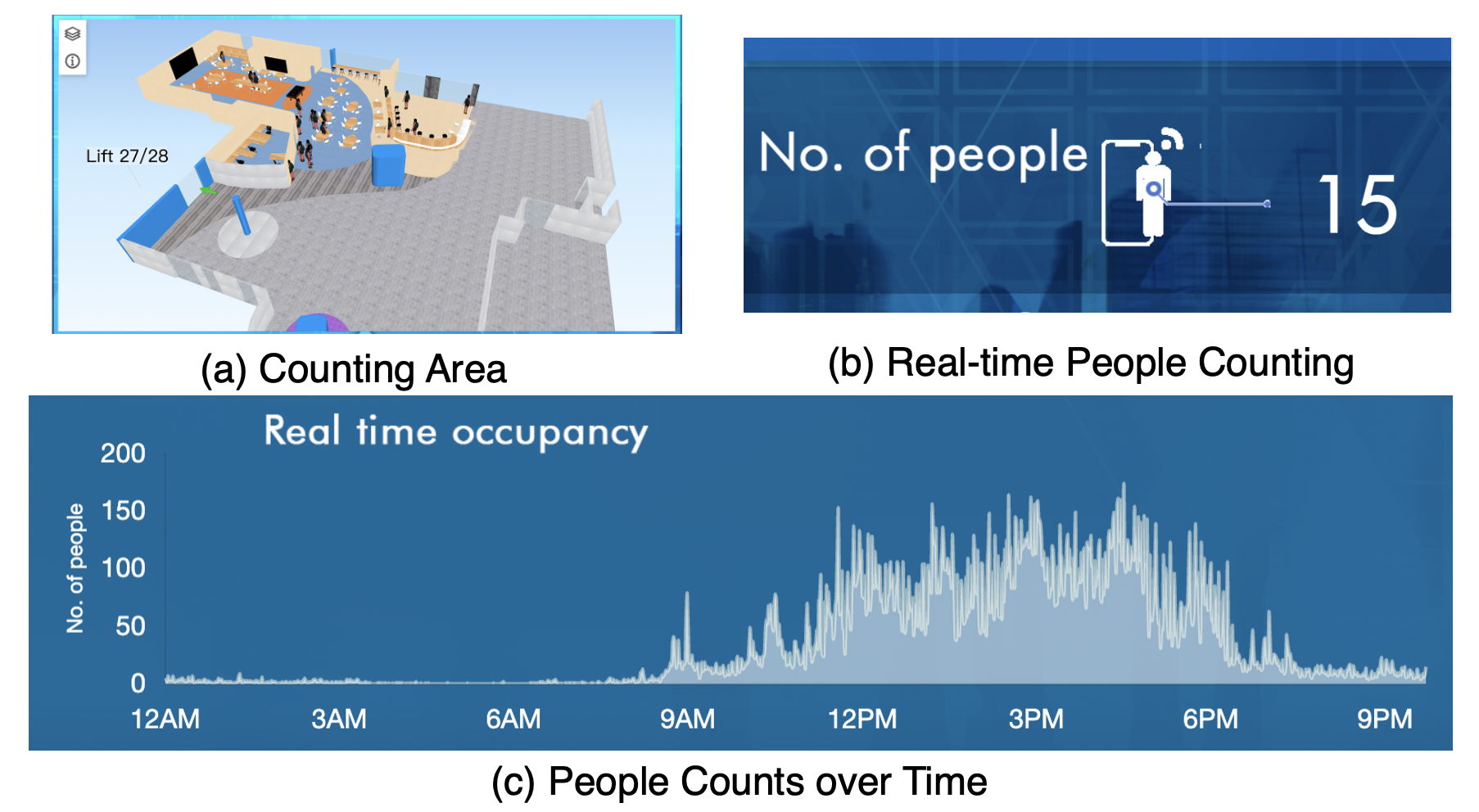}
    \caption{User interface (UI) of the counting system (based on \sysname{}) in a study room. }
    \label{fig:sys_demo}
    \Description{system demo}
\end{figure}

Finally, we demonstrate the user interface (UI) of our existing counting system based on \sysname{} in \FigureWord~\ref{fig:sys_demo}. The demonstrated counting area is a public study room on a campus, whose 3D model is shown in \FigureWord~\ref{fig:sys_demo}(a). In the area, the real-time people counts are estimated using the onsite APs, as shown in \FigureWord~\ref{fig:sys_demo}(b), and the device-to-person ratio is derived from the atrium.   Furthermore, the system visualizes the counting history, as shown in \FigureWord~\ref{fig:sys_demo}(c). The UI has been shown on an LED display and a public website, for use by the university community. 

\RankIIPart{Related Work}
\label{sec:related}

In this section, we review the related works in Wi-Fi-based device counting and the existing sensing technologies for people counting. \\

\noindent
{\bf The Counting of Wi-Fi Devices: }
The simplest method to count Wi-Fi devices is to count the number of devices connected to APs~\cite{10.1145/3448084, 10.1145/3495003, 10.1145/3596237}. While straightforward, the number of Wi-Fi connections does not involve the unconnected devices, leading to large counting errors~\cite{ZhangCrowdTelescopeWiFipositioningbasedmultigrained2023, QuExploringpatronbehavior2023, carroll2023localization, LiSenseFlowExperimentalStudy2015}. To account for the unconnected ones, subsequent works count devices using Wi-Fi probing, which counts the number of unique MAC addresses in PRFs within a window, termed the MAC-counting approach~\cite{oppokhonov2022analysis, determe2022monitoring, YaikMeasuringAccuracyCrowd2016a}. However, this approach is vulnerable to MAC address randomization, where one device may have multiple virtual MAC addresses. Since the randomization process is often difficult to reverse~\cite{furuya2021indoor,StanciuPrivacyfriendlystatisticalcounting2023, vanhoefWhyMACAddress2016, martinStudyMACAddress2017}, state-of-the-art (SOTA) approaches associate PRFs with their broadcast device using PRF messages, which often rely on machine learning~\cite{TorkamandiOnlineMethodEstimating2021, NittiiABACUSWiFiBasedAutomatic2020, 9888045, perez2024randomization}. For example, numerous machine learning models, such as Bayesian networks, decision trees, Gaussian mixture models, and deep learning models, have been studied to associate PRFs based on their messages, including information elements, sequence numbers, received signal strength, and arrival times~\cite{9888045, robyns2017noncooperative, tan2021efficient, matte2016defeating, perez2024randomization, jin2024over}. Although accurate association has been achieved, deploying these association models is inefficient and time-consuming, as they require dedicated efforts to support machine learning, and their association model needs regular professional expertise to accommodate new Wi-Fi devices and models for system maintenance~\cite{dong2021survey, zhou2022domain}. Consequently, SOTA approaches are unsuitable for scenarios that require fast deployment, easy maintenance, or those with restricted computing power. In addition, previous approaches often lack a theoretical understanding of the counting bias. To bridge this gap, we propose \sysname{}, the first learning-free approach to count Wi-Fi devices under MAC address randomization, with theoretically derived counting unbiasedness and error model. \\

\noindent
{\bf People Counting Technologies: }
Several sensing technologies have been considered and used for people counting, such as camera, LiDAR, and mmWave radar~\cite{song2021choose, liu2020adaptive, peng2024single}. 
However, these sensors often have limited sensing range, incur extra cost due to infrastructure setup, and might pose privacy concerns. 
To address these issues, Wi-Fi-based people counting has been explored, which can be categorized into active and passive approaches. 
The active approach counts people by analyzing the physical Wi-Fi signal propagating over the counting area (i.e., channel state information), which often requires specialized sensing devices~\cite{wang2020passive, yang2018wi, jiang2023pa,sobron2018device}. On the other hand, the passive approach listens to the PRFs and studies how to calibrate the number of PRFs to estimate the people count, which, unfortunately, is often inaccurate~\cite{WuCrowdEstimatorApproximatingcrowd2018, 10.1145/3264925, DuivesEnhancingcrowdmonitoring2020, Haofinegrainedcrowdanalysis2023, HuangPedestrianflowestimation2021}. In this paper, we demonstrate that \sysname{}, incorporating the calibration scheme, can significantly improve the accuracy of Wi-Fi-based passive people counting. 

\RankIIPart{Conclusion}
\label{sec:conclusion}

Wi-Fi devices, for network discovery, sporadically broadcast probe request frames (PRFs) with randomized MAC addresses to the nearby access points (APs). 
We study counting the number of Wi-Fi devices within AP coverage by listening to their PRFs over a window. 
While such a well-established research problem has been achieved accurate, previous counting schemes address MAC address randomization by means of using machine learning to associate the PRFs with their broadcast devices. This, however, results in inefficient and time-consuming system setup (training data curation and cleaning, model training, hyperparameter tuning), sophisticated system maintenance (e.g., model adaptation to accommodate new Wi-Fi device brands and models), and high computing requirements for machine learning. 

To lower deployment costs, we propose \sysname{}, an accurate, lightweight, and learning-free approach that counts devices based on the rate at which APs receive PRFs, i.e., probing rate. \sysname{} estimates the device count by a simple and effective closed-form rate model. We have proved that the rate model gives an unbiased estimator of the device count time-averaged over a window and deduced the count variance in terms of the window size, number of PRFs in the window, and the variance of the probing interval. Furthermore, we demonstrate how to extend \sysname{} to people counting via a multimodal calibration scheme based on our experience on an existing counting system. 
We have implemented \sysname{} and conducted extensive experiments on device counting and people counting in various diverse environments with various count levels, specifically, in a lab room, a building entrance, a campus atrium, a bus station, and a study room. The experimental results show that our learning-free \sysname{} can achieve comparable counting accuracy with SOTA learning-based device counting and significantly improve previous Wi-Fi-based people counting by cutting 66\% errors. Additionally, our results show that the error model provides a sound estimation of the counting errors. 

\sysname{} is learning‑free, unbiased, and scales gracefully to crowded scenarios, which are key properties for deployments that demand fast setup and low maintenance. In future work, we plan to further improve its counting accuracy and environmental adaptability. This may be achieved by incorporating additional PRF features %(e.g., information elements) 
and designing adaptive window‑size tuning.

\bibliographystyle{ACM-Reference-Format}
\bibliography{ref} %references
\appendix
\section{Proof of the Unbiasedness of Rate Model}
\label{app: proof}

We provide a proof of $\mathbb{E} (\widehat{N}) = \overline{N}$ below by showing that the whole system is equivalent to a 
\emph{renewal process}. In a renewal process, we consider the summation of a sequence of positive IID random variables $X_{k}$ given as
\begin{equation}
    \label{eq:iid_sum}
    S_K = \sum_{k=1}^{K}X_{k}, 
\end{equation} 
where $\mathbb{E}(X_{k}) = \overline{X}$.
We define a \emph{counting random variable} $L(S)$ as the maximum number of IID random variables ($X_{k}$) 
such that their summation $S_{L(S)} \leq S$. 
Namely, we have $\sum_{k=1}^{L(S)}X_{k} \leq S$ and $\sum_{k=1}^{L(S)+1 }X_{k} > S$.

An important property of the renewal process is the \emph{Elementary Renewal Theorem}, 
which states that
\begin{equation}
    \label{eq:renewal_theorem}
    \frac{\mathbb{E}(L(S))}{S} = \frac{1}{\overline{X}}
\end{equation}
when $S$ is sufficiently large. 

We elaborate that our system is equivalent to a \emph{renewal process} using the single-device example shown in \FigureWord~\ref{fig: probing}. In this figure, the APs received $b_n=2$ bursts separately at $t_1$ and $t_2$, and the device enters the area at time $x$ and leaves at $y$. 
The dwell time of this device is calculated as 
\begin{equation}
    \label{eq:dwell_expect}
        d_n = y - x = \beta_1 \tau_1 + \tau_2 + \beta_3 \tau_3, 
\end{equation} 
where $\beta_1$ and $\beta_3$ represent the fraction of $\tau_1$ and $\tau_3$ with $0 < \beta_1, \beta_3 < 1$. Since the device enters and leaves the area largely independent of the probing, with equal probabilities, $x$ could be any value between $t_0$ and $t_1$, and $y$ could be any value between $t_2$ and $t_3$. Therefore, $\beta_1$ and $\beta_3$ are IID \emph{uniform random variables} with $\mathbb{E}(\beta_1)=\mathbb{E}(\beta_3)=1/2$.

When computing the total dwell time ($D$), we use $T_F$ to denote the set of \emph{fractional} intervals $\{\beta_i \tau_i \}$ 
where only part of $\tau_i$ is summed up (e.g., $\beta_1 \tau_1$ and $\beta_3 \tau_3$), 
and $T_I$ to denote the set of \emph{integral} intervals $\{\tau_i\}$ 
where the whole of $\tau_i$ is summed up (e.g., $\tau_2$). 
It is important to note that both $T_F$ and $T_I$ are sets of IID random variables
where $\mathbb{E}(\beta_i \tau_i) = \overline{\tau}/2$ and $\mathbb{E}(\tau_i) = \overline{\tau}$.

Generally speaking, $\mathbf{N}$ devices with $B$ probing bursts would have $2\mathbf{N}$ fractional intervals and $B-\mathbf{N}$ integral intervals.
Therefore, the dwell time of all the devices can be written as
\begin{equation}
    \label{eq:dwell_total}
        D = \sum_{\beta_i \tau_i \in T_F} \beta_i \tau_i + \sum_{\tau_i \in T_I} \tau_i.
\end{equation}

We let $D_F = \sum \beta_i \tau_i$ for $\beta_i \tau_i \in T_F$ and 
$D_I = \sum \tau_i$ for $\tau_i \in T_I$.
Both $D_F$ and $D_I$ are the sums of IID random variables similar to $S_K$ in \EquationWord(\ref{eq:iid_sum}). 
Clearly, we have $D = D_F + D_I$. By applying the Elementary Renewal Theorem given in \EquationWord(\ref{eq:renewal_theorem}), we have
\begin{equation}
    \label{eq:renewal_theorem_I}
    \frac{\mathbb{E}(B-\mathbf{N})}{D_I} = \frac{1}{\overline{\tau}}
\end{equation}
and
\begin{equation}
    \label{eq:renewal_theorem_F}
    \frac{\mathbb{E}(2\mathbf{N})}{D_F} = \frac{2}{\overline{\tau}}
\end{equation}
when $D_F$ and $D_I$ are sufficiently large.

Given \EquationWord(\ref{eq: n_hat}), we also have
\begin{equation}
    \begin{aligned}
    \label{eq:estimator_1}
    \mathbb{E} (\widehat{N}) & = \mathbb{E}\left(\frac{R}{\overline{r}}\right) = \mathbb{E}\left(\frac{B\overline{\tau}}{w}\right) 
    = \frac{\overline{\tau}}{w}\mathbb{E}(B) \\
      & = \frac{\overline{\tau}}{w} \left(\mathbb{E}(B-\mathbf{N}) + \frac{1}{2}\mathbb{E}(2\mathbf{N})\right).
    \end{aligned}
\end{equation}
Based on \EquationWords(\ref{eq: mean_device}), (\ref{eq:renewal_theorem_I}), (\ref{eq:renewal_theorem_F}), and (\ref{eq:estimator_1}), 
when $\overline{N}$ and $w$ are sufficiently large so that $D$ is also sufficiently large, we have
\begin{equation}
    \label{eq:estimator_2}
    \mathbb{E} (\widehat{N}) = \frac{\overline{\tau}}{w} \left(\frac{D_I}{\overline{\tau}} 
     + \frac{1}{2} \frac{2 D_F}{\overline{\tau}}\right) = \frac{D}{w} = \overline{N}.
\end{equation}
This shows that the rate model is an unbiased estimator of the window-averaged device count,  
which ends the proof.

\end{document}